\documentclass[11pt,epsf,amstex]{article}
\usepackage [dvips]{graphicx}
\usepackage{amsmath}
\usepackage{times}
\usepackage{psfig}
\usepackage{epsfig}
\usepackage{amssymb}

\newcommand{\dd}{\text{d}}
\newcommand{\sR}{\text{\tiny R}}
\newcommand{\ee}{\text{e}}

\newcommand{\p}{\partial}

\newcommand{\br}{\text{\bf r}}

\newcommand{\eps}{\varepsilon}

\newcommand{\bo}{\text{\bf 0}}
\newcommand{\bq}{\text{\bf q}}

\begin{document}
\newenvironment{dessins}{{\bf Figures}}{}
\begin{center}
{\huge{\bf Fluctuation-induced first order\\
transition in a nonequilibrium\\[2mm]
steady state}}
\end{center}
\vspace{2cm}
\begin{center}
{Klaus Oerding$^a$, Fr\'ed\'eric van Wijland$^b$, Jean-Pierre Leroy$^b$ and
Hendrik Jan Hilhorst$^b$}
\end{center}

\noindent {\small $^a$ Institut f\"ur Theoretische Physik III, Heinrich Heine
Universit\"at, 40225 D\"usseldorf, Germany.\\

\noindent $^b$ Laboratoire de Physique Th\'eorique, Universit\'e de Paris-Sud,
91405 Orsay cedex, France.}\\\\

\begin{center}{\bf Abstract}\\
\end{center}
{\small We present the first example of a phase transition in a nonequilibrium
steady-state that can be argued analytically to be first order. 
The system of interest is a two-species reaction--diffusion problem
whose control parameter is the total density $\rho$.
Mean-field theory predicts a second-order
transition between two stationary states 
at a critical density $\rho=\rho_c$.
We develop a phenomenological picture that, instead,
below the upper critical dimension $d_c=4$, predicts 
a first-order transition. This picture is confirmed 
by hysteresis found in numerical simulations,
and by the study of a renormalization-group
improved equation of state. The latter approach is inspired by the
Weinberg-Coleman mechanism in QED. }

\vskip 2cm

Pr\'epublication L.P.T. Orsay 99/46.
\newpage
\section{Motivation}
\subsection{Fluctuation induced first order transitions}
In the realm of equilibrium critical phenomena it is well-known that
systems which in high space dimension $d$ undergo
a first-order transition, 
may exhibit a second-order transition below their upper critical
dimension $d=d_c$.
Examples are spin systems with cubic
anisotropy~\cite{Domany,Rudnick,AY}, type-II
superconductors~\cite{Chen}, and the three-state Potts model~\cite{Wu}.
In equilibrium phenomena
the phase diagram can be deduced from the analysis of the global extrema of a 
free
energy functional. The global free energy
minima correspond to stable
phases. For $d<d_c$ this functional has to incorporate fluctuation
effects. When fluctuations modify the energy
landscape to the point of changing a second-order transition into a
first-order one, one has a {\it fluctuation-induced} first-order
transition. This 
phenomenon is
also said to be due to the Coleman-Weinberg mechanism~\cite{ColemanWeinberg}. 
Indeed formally similar
phenomena were first found in the study of the coupling of QED radiative
corrections to a charged scalar field.\\

In nonequilibrium systems there is no such concept as a free energy;
the steady state phase diagram cannot be 
deduced from a thermodynamic potential, and its statistical mechanics is
not based on a partition function. The starting point, instead, is usually 
an evolution equation (often a master equation),
whose stationary solutions are to be determined and yield the
steady-state phase diagram. This indirect definition of the phase
diagram renders analytic approaches very
cumbersome. In the past twenty years techniques have
been devised to find the steady states of such master equations and
extract from them physical properties
of interest
(order parameter, correlation functions,$\ldots$). 
These techniques include
short time series expansion, numerical simulations, real-space renormalization
and field-theoretic approaches. 
Nonequilibrium steady states (NESS) may undergo
phase transitions in the same way as do equilibrium states.
Several examples of {\it continuous} transitions in such systems
have been found and well studied. 
To our knowledge, the only {\it first-order} transitions in NESS 
known today occur in the asymmetric exclusion
model in one space
dimension, as demonstrated analytically in Ref.~\cite{Derrida}.
This model belongs to the subclass of {\it driven diffusive systems:} due to
an externally applied field these systems have a spatially anisotropic
current carrying NESS. The
occurrence of first order
transitions in certain other driven diffusive systems~\cite{KornissZiaBeate}
is also suggested by
numerical simulations. 
Finally, Schmittmann and Janssen~\cite{BeateJanssen}
have argued field-theoretically
that a similar fluctuation mechanism may induce a first-order
longitudinal transition (high and low density stripes
perpendicular to the driving field) in a driven diffusive system with a
single conserved density.\ \

The present study bears on the phase transition in the
NESS of a diffusion-limited reaction between two species of particles.
This  system does {\it not} belong to the subclass of
driven systems: it remains spatially isotropic under all circumstances.\\

We exhibit here, for the first time with analytic arguments, a 
{\it fluctuation-induced} first order
transition in the steady-state of a reaction-diffusion process. We
present an analytic procedure  that
allows to access the phase diagram of the system in a very explicit  fashion.
The outline of the article is as follows. In the next subsection we
define the two-species reaction-diffusion model. In section 2 
we recall
some known properties of its phase diagram. We also introduce 
the field-theoretical formalism, which will be our main
tool of analysis. Section 3 presents a heuristic
approach to the first-order transition in terms of a nucleating and diffusing
droplet picture. In section 4 we report on
numerical simulations on a two-dimensional system, in which a
hysteresis loop is observed for the order parameter. This confirms the
suspected existence of a first order transition below the critical
dimension $d_c=4$.
Sections 5 through 8 contain the field-theoretic approach: the
derivation of a renormalization group improved equation of state, valid
in dimension $d=d_c-\varepsilon$,
followed by the study of its
solutions and of their stability with respect to spatial perturbations. We
conclude with a series of possible applications.

\subsection{Reaction-diffusion model}
Particles of two species, $A$ and $B$, diffuse in a $d$-dimensional space with 
diffusion
constants $D_A$ and $D_B$, respectively. Upon encounter an $A$ and a $B$ are 
converted
into two $B$'s at a rate $k_0$ per unit of volume,
\begin{equation}\label{contamination}
A+B\stackrel{k_0}{\rightarrow} B+B
\end{equation}
Besides a $B$ spontaneously decays into an $A$ at a rate
$\,\gamma$,
\begin{equation}\label{guerison}
B\stackrel{\gamma}{\rightarrow}A
\end{equation}
Denoting 
the local $A$ and $B$ densities by $\rho_A$ and $\rho_B$, respectively,
we can write the mean-field equations as 
\begin{eqnarray}
\label{eqchampmoyen}
\p_t\rho_B=D_B\Delta\rho_B-\gamma\rho_B+k_0\rho_A\rho_B\nonumber\\
\p_t\rho_A=D_A\Delta\rho_A+\gamma\rho_B-k_0\rho_A\rho_B
\end{eqnarray}
The total particle density $\rho$, 
is a conserved quantity and will be the control
parameter. In the initial state particles are distributed randomly and 
independently,
with a given fraction of each species. 
Let $\rho_A^{\text{st}}$ and $\rho_B^{\text{st}}$
be the steady state values of the $A$ and $B$ particle density,
respectively. Obviously their sum is equal to $\rho$.
One easily derives from (\ref{eqchampmoyen}) that there exists a
threshold density $\rho_c=\gamma/k$
such that for $\rho>\rho_c$ the steady state of the 
system
is {\it active}, that is, has $\rho_B^{\text{st}}>0$, 
and that for $\rho<\rho_c$ it is
absorbing, that is, has $\rho_B^{\text{st}}<0$.
Hence $\rho_B^{\text{st}}$ is the order parameter for this system, and
an important question is how this quantity behaves at the transition point.

This model may be cast into the form of a field theory (and then turns
out to generalize the field theory of the Directed Percolation problem). 
It was shown by field-theoretic methods in
\cite{KSS} and \cite{WOH} 
that for $0<D_A\leq D_B$
the transition between the steady
states at $\rho=\rho_c$ is continuous. It is characterized by a set of critical exponents 
that differ from
their mean-field values in $d<d_c=4$. 
Whereas for $D_A\leq D_B$ the phase diagram may be obtained by
convential tools of analysis, such as renormalization group approaches based on
a field-theoretic formulation of the dynamics, the case 
$D_A>D_B$ cannot be analyzed along the same lines. In technical terms, the
renormalization group flows away to a region where the theory is ill-defined.
We interpret this as meaning
that there is no continuous transition, and a natural idea,
if one still believes in the existence of a transition, is the occurrence of
a first-order one. The article is concerned with the case
$D_A>D_B$.
\section{Field-theoretic formulation}
\subsection{Langevin equations}
There are at least two different ways to construct a field theory that
describes a reaction-diffusion problem such as the one we just defined.
One of them is to encode
the stochastic rules for particle diffusion and reaction in a 
master equation for the probability of occurrence of a state of given local
particle numbers at a given time. The master equation may be converted
into an exactly equivalent field theory 
following methods that were pioneered by Peliti~\cite{Peliti} and
others.
We will follow a different way of proceeding that has been  widely
employed, in particular, by Janssen and co-workers~\cite{J81}.
We postulate for the space and time dependent
densities of the $A$ and $B$ particles 
two Langevin equations in which the noise terms 
have been deduced by heuristic
considerations. The result of this 
approach differs from Peliti's up to terms
that are irrelevant in the limit of large times and distances.\\

We switch now to the notation of field theory and denote
by $\psi(\br,t)$ the coarse-grained $B$ particle density and by 
$m(\br,t)$ the deviation from average of the
coarse-grained total particle density. The deterministic part of the Langevin
equations to be constructed should 
be the conventional mean-field
reaction-diffusion PDE's of equation (\ref{eqchampmoyen}).
Upon adding two noise terms $\eta$ and $\xi$ we get
in the new notation, and after redefinition of several parameters,
\begin{equation}\label{Lan1}
\p_t\psi=\lambda\Delta\psi-\lambda\tau\psi-\frac{\lambda g}{2}\psi^2-\lambda f
m\psi+\eta
\end{equation}
\begin{equation}\label{Lan2}
\p_t m=\Delta m+\lambda\sigma\Delta\psi+\xi
\end{equation}
Here $\lambda=D_B/D_A$ is the ratio of the diffusion constants; 
$\tau$ is proportional to the
deviation of the total density from its mean-field critical value
$\gamma/k$; $g, \tilde{g}$ and $f$ are proportional to the contamination rate
$k$;
and the parameter $\sigma$, which will play a key role in the present
study, is proportional to $1-\frac{D_B}{D_A}$. 

For mathematical convenience we will want $\eta$ and $\xi$ to represent
mutually uncorrelated Gaussian white noise.
The noise $\eta$ in the
equation for $\psi$ should vanish with $\psi$ since $\psi=0$ is an absorbing
state. For the autocorrelation of $\eta$ we therefore 
retain the first term of a hypothetical series expansion in
powers of $\psi$. This procedure is best described in \cite{J81,KSS}. The
autocorrelation of $\xi$ should be such that $m$ is locally conserved.
With these conditions
the simplest possible expressions for
the autocorrelations of $\eta$ and $\xi$ are, explicitly,
\begin{eqnarray}
\langle\eta(\br,t)\eta(\br',t')\rangle=
\lambda\tilde{g}\psi(\br,t)\delta^{(d)}(\br-\br')\delta(t-t')
\nonumber\\
\langle\xi(\br,t)\xi(\br',t')\rangle=
2\nabla_{\br}\nabla_{\br'}
\delta^{(d)}(\br-\br')\delta(t-t')
\end{eqnarray}
The Langevin equation (\ref{Lan1}) is to be understood
with the It\^o discretization rule. It is also possible to derive these
equations {\it ab initio} by the operator formalism used in \cite{WOH}. \\

Using the Janssen-De Dominicis formalism~\cite{J76,DeD} 
we obtain the physical observables
as functional integrals over four fields
$\bar{\psi},\psi,\bar{m},m$ weighted by a factor
$\exp(-S[\bar{\psi},\psi,\bar{m},m])$, with
\begin{equation}\label{actioncomplete}\begin{split}
S[\bar{\psi},\psi,\bar{m},m]=&\int\dd^dx\dd
t\;\Big[\bar{\psi}(\p_t+\lambda(\tau-\Delta))\psi+\bar{m}(\p_t-\Delta
)m\\&-(\nabla\bar{m})^2-\lambda\sigma\bar{m}\Delta\psi+
\frac{\lambda
g}{2}\psi^2\bar{\psi}-\frac{\lambda\tilde{g}}{2}\psi\bar{\psi}^2+\lambda
f\bar{\psi}\psi m\Big]
\end{split}\end{equation}
It is possible (and sometimes more practical) to eliminate the fluctuating
density field $m$ and its response field $\bar{m}$, which yields an
effective action for the $\bar{\psi},\psi$ fields alone.

\subsection{Mean-field equation of state}
Our starting point is the
action describing the dynamics of the system in the presence of an arbitrary 
source
of $B$ particles $\lambda h(\br,t)$ in which the fluctuating density $m$ and its
response field $\bar{m}$ have been integrated out. It reads
\begin{equation}\label{Actioneffpourpsi}\begin{split}
S[\bar{\psi},\psi]=&\int\dd^d r\dd
t\;\Big[\bar{\psi}(\p_t+\lambda(\tau-\Delta))\psi+\frac{\lambda
g}{2}\psi^2\bar{\psi}-\frac{\lambda\tilde{g}}{2}\psi\bar{\psi}^2-\lambda 
h\bar{\psi}\Big]\\&
-\int\dd^d r\dd^dr'\dd t\dd t'\;\Big[\frac{(\lambda
f)^2}{2}\bar{\psi}\psi(\br,t)
C_0(\br-\br';t-t')\bar{\psi}\psi(\br',t')\\&-\lambda^2\sigma
f\bar{\psi}\psi(\br,t)G_0(\br-\br';t-t')\Delta\psi(\br',t')\Big]
\end{split}\end{equation}
where the spatial Fourier transforms of $G_0$ and $C_0$ are
\begin{equation}
G_0(\bq;t-t')=\Theta(t-t')\ee^{-\bq^2(t-t')},\;\;\;
C_0(\bq;t-t')=\ee^{-\bq^2|t'-t|}
\end{equation}
We first look for an {\it a priori} inhomogeneous steady state (in terms of
Fourier transforms, one takes the
limit $\omega\rightarrow 0$) then one specializes the study to a homogeneous
steady state (and one takes the limit $\bq\rightarrow \bo$). The limit of
infinite times (corresponding to a system reaching a steady-state) and the limit 
of a homogeneous system do not commute. 
In the limit of a vanishing source term 
the mean-field equation of state for a homogeneous order parameter $\Psi$ is
found by imposing that
\begin{equation}
\lim_{\bq\rightarrow \bo}\lim_{\omega\rightarrow 0}\frac{\delta S}{\delta
\bar{\psi}(\bq,\omega)}[0,\Psi]=\lambda\Psi\left({\tau}+\frac{\bar{g}}{2}\Psi\right)=0
\end{equation}
with $\bar{g}\equiv g-2\lambda \sigma f$. It is important to note that written
in terms of the original parameters $\bar{g}>0$ as long as $\frac{D_B}{D_A}>0$
(the cases
$D_B=0$ or $D_A=0$ would require a separate study). One may conclude that
the steady state is active (a finite fraction of $B$'s survive indefinitely) if
$\tau>0$, while the system eventually falls into an absorbing $B$-free state if
$\tau<0$. Mean-field therefore predicts a {\it continuous} transition between 
those states at
$\tau=0$, independently of the ratio of the diffusion
constants.

\subsection{Renormalization}
In order to go beyond mean-field we have performed a one-loop perturbation 
expansion
of the two and three-leg vertex functions. Renormalization is then required to
extract physically relevant information from this expansion. We shall proceed 
within
the framework of dimensional regularization and of the minimal subtraction 
scheme. In
order to absorb the $\eps$-poles in the vertex functions into a 
reparametrization
of coupling constants and fields we introduce the renormalized
quantities $\psi_{\sR}$, $\lambda_{\sR}$, $\rho_{\sR}$ etc. defined by
\begin{eqnarray}\label{Rcouplings}
\bar{\psi} = Z_{\bar{\psi}}^{1/2} \bar{\psi}_{\sR} \qquad
\psi = Z_{\psi}^{1/2} \psi_{\sR} \qquad Z = (Z_{\bar{\psi}}
Z_{\psi})^{1/2} \nonumber \\
Z \lambda = Z_{\lambda} \lambda_{\sR} \qquad \lambda \rho =
\lambda_{\sR} \rho_{\sR} \qquad Z_{\psi}^{1/2} \lambda \sigma
= \lambda_{\sR} \sigma_{\sR} \nonumber \\
Z_{\lambda} \tau = Z_{\tau} \tau \qquad
A_{\eps}^{1/2} Z_{\lambda} f = Z_{f} f_{\sR} \mu^{\eps/2}
\qquad Z_{\lambda} h = Z_{\psi}^{1/2} h_{\sR} \\
A_{\eps}^{1/2} Z_{\lambda} Z_{\psi}^{1/2} g =
\sigma_{\sR} \mu^{\eps/2} (Z_{g} g_{\sR} + W f_{\sR}) \qquad
 A_{\eps}^{1/2} Z_{\lambda} Z_{\bar{\psi}}^{1/2}
\sigma_{\sR} \tilde{g} = Z_{\tilde{g}} \tilde{g}_{\sR}
\mu^{\eps/2} . \nonumber
\end{eqnarray}
Here $\mu$ denotes an external momentum scale. The renormalization
factors depend on $u = \tilde{g}_{\sR} g_{\sR}$, $v = f_{\sR}^{2}$ and
$w = f_{\sR} \tilde{g}_{\sR}$ and are at one-loop order given by
\begin{eqnarray}\label{Zfactors}
Z = 1 + \frac{u}{4 \eps}  - \frac{2 v}{\eps (1+\rho)^{2}}
- \frac{3+\rho}{2 \eps (1+\rho)^{2}} w \\
Z_{\lambda} = 1 + \frac{u}{8\eps} - \frac{2 \rho v}{\eps
(1+\rho)^{3}} - \frac{\rho^{2}+4\rho-1}{4\eps (1+\rho)^{3}}
w \\
Z_{\tilde{g}} = 1 + \frac{u}{\eps} - \frac{2(3+\rho)}{
\eps (1+\rho)^{2}} v - \frac{2(2+\rho)}{\eps (1+\rho)^{2}}
w \\
Z_{g} = 1 + \frac{u}{\eps} - \frac{2(3+\rho)}{\eps
(1+\rho)^{2}} v - \frac{5+3\rho}{\eps (1+\rho)^{2}} w \\
Z_{f} = 1 + \frac{u}{2\eps} - \frac{2 v}{\eps (1+\rho)^{2}}
- \frac{2+\rho}{\eps (1+\rho)^{2}} w \\
W = \frac{4v}{\eps \rho (1+\rho)} + \frac{2w}{\eps \rho
(1+\rho)} .
\end{eqnarray}
Since only $Z$ is fixed by the renormalization conditions
but not the individual factors $Z_{\bar{\psi}}$ and $Z_{\psi}$
we may set $Z_{\psi} = 1$.

\subsection{Renormalization group and fixed points}
\begin{figure}[h]
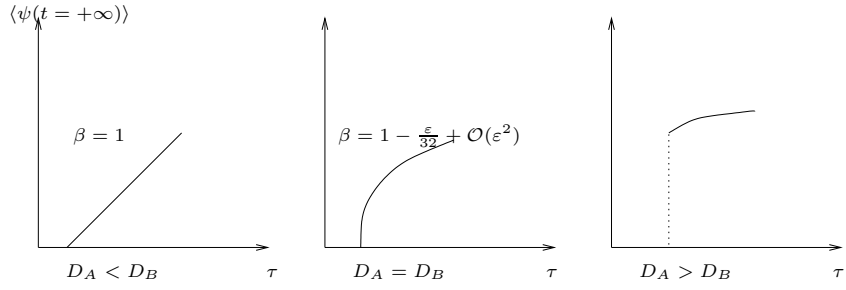
\caption{Phase diagram in the $(\tau,\psi(t=\infty))$ plane 
(the ordinate is the
steady-state density of $B$ particles) for $\lambda>1$,
$\lambda=1$ and a conjecture for $\lambda<1$. Also shown is the order parameter
exponent $\beta$.}
\label{phasediag}
\end{figure}

>From the above $Z$-factors and the definition of the renormalized couplings one
finds the flow equations for the renormalized couplings. These read
\begin{eqnarray}
  \beta_{u} = \mu \left.\frac{\dd u}{\dd \mu}\right|_{\rm bare} =
u \left(-\eps + \frac{3 u}{2} - \frac{2(5+5\rho+2\rho^{2}) v}{(
1+ \rho)^{3}} - \frac{2(4+5\rho+2\rho^{2})w}{(1+\rho)^{3}} \right)
\nonumber \\
+ w \left(\frac{4 v}{\rho (1+\rho)} + \frac{2 w}{\rho (1+\rho)} 
\right) \label{betu} \\
  \beta_{v} = \mu \left.\frac{\dd v}{\dd \mu}\right|_{\rm bare} =
v(-\eps + 2\kappa) =
v \left(-\eps + \frac{3 u}{4} - \frac{4 v}{(1+\rho)^{3}}
- \frac{(9+8\rho+3 \rho^{2})w}{2 (1+\rho)^{3}} \right) \\
  \beta_{w} = \mu \left.\frac{\dd w}{\dd \mu}\right|_{\rm bare} =
w \left(-\eps + \frac{9 u}{8} - \frac{2(4+2\rho+\rho^{2}) v}{
(1+\rho)^{3}} - \frac{3(9+8\rho+3\rho^{2})w}{4(1+\rho)^{3}}
\right) \\
  \beta_{\rho} = \mu \left.\frac{\dd \rho}{\dd \mu}\right|_{\rm
bare} = -\zeta \rho =
\rho \left( \frac{u}{8} - \frac{2 v}{(1+\rho)^{3}} -
\frac{(7+4\rho+\rho^{2}) w}{4 (1+\rho)^{3}} \right)
\end{eqnarray}
and the Wilson function is
\begin{equation}
\gamma = \mu \left.\frac{\dd \ln Z}{\dd \mu}\right|_{\rm bare} =
- \frac{u}{4} + \frac{2 v}{(1+\rho)^{2}} + \frac{(3+\rho)w}{2
(1+\rho)^{2}} \label{gam} 
\end{equation}
The combination $\lambda \rho$ remains equal to 1 along the renormalization 
flow.
In equations~(\ref{betu})-(\ref{gam}) all $\mu$-derivatives are at 
fixed bare parameters. The renormalization group flow has three nontrivial fixed 
points:
the well-known directed percolation fixed point with $v=w=0$ and
$u = u_{\text{DP}} = 2\eps/3$, the symmetric ($w=0$) fixed point
$(u_{s}, v_{s}, \rho_{s}) = (2\eps, 27\eps/64, 1/2)$
and the asymmetric fixed point $(u_{a}, v_{a}, w_{a}, \rho_{a})$
with
\begin{eqnarray}
u_{a} = \frac{4\eps}{2+\rho_{a}} \qquad
v_{a} = \frac{1+\rho_{a}}{4} \eps \qquad
w_{a} = - \frac{1-5\rho_{a}}{\rho_{a}} \eps \nonumber \\
\rho_{a} = (2+\sqrt{3})^{1/3} + (2-\sqrt{3})^{1/3} - 2
\end{eqnarray}
at leading order in $\eps$. The continuous phase transitions
described by these fixed points have already been studied in other
publications~\cite{KSS,WOH}. The symmetric fixed point ($w=0$) is
unstable with respect to the variable $w$. It corresponds to the
case of equal diffusion constants $D_{A}=D_{B}$, whereas the asymmetric
fixed point with $w<0$ governs the critical behavior for
$D_{A} < D_{B}$. Since the sign of $w$ is conserved along the
renormalization group flow the asymmetric fixed point cannot be reached
for $w>0$. Therefore there is no fixed point for $w>0$
($D_{A} > D_{B}$). In order to study the phase transition for
$D_{A} > D_{B}$, which is the regime of interest, we consider in the next 
sections the solutions of the renormalization group flow in more detail. Figure 
\ref{phasediag} shows a
plot of the steady-state density of $B$'s as a function of $\tau$ (the deviation
of the total density with respect to its mean-field critical value), for
$\lambda =1$ and $\lambda>1$.

Assuming the existence of a first-order transition for $\lambda<1$ (which would
make of $\lambda=1$ a tricritical point), the jump of
the order parameter across the transition point is related to the properties of
the symmetric fixed point and should scale according to the
tricritical scaling predictions developed by Lawrie and Sarbach~\cite{Lawrie},
\begin{equation}
\rho_B(\tau_c^-)-\rho_B(\tau_c^+)\propto \sigma^{1/\delta},\qquad
\delta=-\frac{\gamma_{s}}{d+\gamma_s}
\end{equation}
where $\gamma_{s}$ is the Wilson function $\gamma$ evaluated at the symmetric
fixed point.

\section{A phenomenological theory}
\subsection{What happens when $D_B<D_A$?}
In dimension $d<4$ the renormalization group flow has no stable fixed point
at finite coupling constants. Nevertheless, we still
expect a phase transition. Here follows a heuristic argument leading to
the conclusion that this is a first order transition. It is based,
essentially, upon adding to the mean-field equations Eqs.~(\ref{eqchampmoyen})
in an approximate
way the fluctuations in the $A$ particle density. Several steps in the argument
are open to criticism but we expect it to provide
the right qualitative picture.\\

Let us consider the system at total particle density $\rho$ and write 
\begin{equation}
\rho=\rho_c+\rho_0
\end{equation}
where $\rho_c$ is the critical density. The mean-field values of the
stationary $A$ and $B$ densities are $\rho_A^{\text{mf}}=\rho_c$ and
$\rho_B^{\text{mf}}=\rho_0$, respectively. 

We imagine the system divided into
regions ("blocks") of volume $L^d$, where $L$ is arbitrary.
Consider a particular block. The instantaneous density in this
block is a fluctuating variable that we denote by
\begin{equation}\label{defrho0}
\rho_L=\rho_c+\rho_0+\delta\rho_L
\end{equation}
where $\delta\rho_L$ is a random term of average zero.\\

We present the argument for the case 
$\rho_0\ll\rho_c$, {\it i.e.} the average $B$ density is much smaller
than the average $A$ density. Then the fluctuations of the total density
are practically identical to those of $\rho_A$. We have in particular 
$\langle\delta\rho_L^2\rangle = \rho_c L^{-d}$, so that the probability
distribution of $\delta\rho_L$ is
\begin{equation}
P(\delta\rho_L)= C\exp\Big(-\frac{L^d\delta\rho_L^2}{2\rho_c}\Big)
\end{equation}
A density fluctuation $\delta\rho_L$ will relax to zero diffusively,
hence on a time scale 
\begin{equation}
T_{\text{fl},L}\sim \frac{L^2}{D_A}
\label{Tfl}
\end{equation}
We are now interested in fluctuations of $\rho_L$ well 
below the critical density ("negative fluctuations"), 
say less than  $\rho_c-\rho_1$. We have
\begin{equation}
\text{Prob}(\rho_L<\rho_c-\rho_1)\sim
\exp\Big(-\frac{L^d(\rho_0+\rho_1)^2}{2\rho_c}\Big)
\label{prob}
\end{equation}
Such a fluctuation will still have the decay time $T_{\text{fl},L}$ given by
(\ref{Tfl}) and therefore stay negative during a time 
\begin{equation}
T_{\text{neg},L}\sim \frac{\rho_1}{\rho_0+\rho_1}\frac{L^2}{D_A}, 
\label{Tneg}
\end{equation}
In the meanwhile the local density of $B$ particles will tend to zero 
with a relaxation time $T_{\text{rel},L}$ which, according to the mean field
equations, in the absence of $B$ diffusion is given by 
\begin{equation}
T_{\text{rel},L}\sim\frac{1}{k\rho_1}
\label{Trel}
\end{equation} 
If $\rho_1$ is so large that $T_{\text{rel},L}\lesssim T_{\text{neg},L}$, 
then during the lifetime of the
negative density fluctuation the $B$ particles will become locally extinct.
Upon combining (\ref{Tneg}) and (\ref{Trel}) we find the condition for
such a $B$ extinguishing density fluctuation. We now take $\rho_1$
exactly large enough for this condition to be 
satisfied, but not larger, since we want to take into account {\it all}
extinguishing fluctuations. This leads to a relation between $\rho_1$ and
$L$, {\it viz.}
\begin{equation}
\frac{\rho_1^2}{\rho_0+\rho_1} = \frac{D_A}{kL^2}
\label{rel1L}
\end{equation}
We now ask what the typical time interval $T_{\text{int},L}$ 
is between two such fluctuations in the same block.
A rough estimate can be made as follows.
The fraction $f_-$ of time spent by the fluctuating density $\rho_L$ below the
value $\rho_c-\rho_1$ is equal to
$f_-\equiv\text{Prob}(\rho_L<\rho_c-\rho_1)$, hence given by
(\ref{prob}). This fraction is composed of short intervals of typical
length $T_{\text{neg},L}$ given by (\ref{Tneg}). The short intervals are
separated by long ones of typical length $T_{\text{int},L}$ that make up
for the remaining fraction, $1-f_-$, of time. Hence 
$T_{\text{neg},L}/T_{\text{int},L}=f_-/(1-f_-)$. Using (\ref{prob}) and
(\ref{Tneg}) we find 
\begin{equation}
T_{\text{int},L}\sim \frac{L^2}{D_A}\exp\Big(\frac{L^d(\rho_1+\rho_0)^2}
{2\rho}\Big)
\label{Tint}
\end{equation}
where our replacing the prefactor $\rho_1/(\rho_0+\rho_1)$ is without
consequences for the remainder of the argument.
The quantity $T_{\text{int},L}$ is the decay time of the $B$ population 
due to density fluctuations on scale $L$, in the absence of $B$ diffusion.
We now take into account the effect of this diffusion. 
The time $T_{B,L}$ needed
for a $B$ particle to diffuse over a distance of order $L$ is
\begin{equation}
T_{B,L}\sim \frac{L^2}{D_B}
\label{TBL}
\end{equation}
All $B$ particles will be eliminated from the system by negative density
fluctuations on scale $L$ unless $T_{B,L} \lesssim T_{\text{int},L}$.
By comparing (\ref{Tint}) and (\ref{TBL}) we obtain for the existence 
of a stationary state with a nonzero $B$ density the condition
\begin{equation}
f(L;\rho_0)\equiv\frac{L^d(\rho_1+\rho_0)^2}{2\rho_c} \gtrsim
\ln\frac{D_A}{D_B}  \qquad \text{for all ~} L\geq a 
\label{cond}
\end{equation}
where $a$ is the lattice parameter and with $\rho_1$ related to $L$ 
by (\ref{rel1L}). 
The key point is now that when $D_B<D_A$, the inequality (\ref{cond})
can be satisfied only for
$\rho_0$ above some threshold $\rho_{0c}$ to be determined below.
Therefore 
\begin{equation}
\rho'_c= \rho_c+\rho_{0c}
\end{equation} 
is the new critical density. 
Since after having survived a  negative density
fluctuation any local $B$ particle density will rapidly return to its average
value $\rho_0$, there is at the new critical density a jump  
in $\rho_B^{\text{st}}$ equal to 
\begin{equation}
\Delta\rho_B^{\text{st}}=\rho_{0c}
\end{equation}

We now determine the threshold value $\rho_{0c}$.
Since the inequality 
(\ref{cond}) has to hold for all $L$, we first determine the
minimum value of its LHS as a function of $L$. In practice the calculation
is most easily done by using $\rho_1$ instead of $L$ as the independent
variable. The minimum occurs for $L=\xi_{\text{min}}$ with
\begin{equation}
\xi_{\text{min}}^2=(2d)^{-2}(4+d)(4-d)\,\frac{D_A}{k\rho_0}
\end{equation}
The values of $\rho_1$ and $f(L;\rho_0)$ at $L=\xi_{\text{min}}$ are
\begin{eqnarray}
\rho_{1,\text{min}}&=&\frac{2d}{4-d}\,\rho_0\\
f(\xi_{\text{min}};\rho_0)&=&(2\rho_c)^{-1}C_d(4-d)^{-2+d/2}
\Big(\frac{D_A}{k}\Big)^{d/2}\rho_0^{2-d/2}
\label{fmin}
\end{eqnarray}
where $C_d=(2d)^{-d}(4+d)^{2+d/2}$. 
The condition $\xi_{\text{min}}\geq a$ leads to
\begin{equation}
(4-d)\frac{D_A}{\rho_0 ka^2}\,\geq\, 1
\label{Lmingeqa}
\end{equation}
and can always be satisfied by choosing $a$ small enough.
Upon inserting (\ref{fmin}) in (\ref{cond}) we find the critical value
$\rho_{0c}$ below which there cannot exist a phase with $B$ particles:
\begin{equation}
\rho_{0c}=\text{cst.}\,\rho_c^{\tfrac{2}{4-d}}\left(\frac{D_A}{k}
\right)^{-\tfrac{d}{4-d}}
(4-d)\ln^{\tfrac{2}{4-d}}\frac{D_A}{D_B}
\label{rho0c}
\end{equation}
Consistency requires that (\ref{Lmingeqa}) be satisfied when for
$\rho_0$ we substitute $\rho_{0c}$ taken from (\ref{rho0c}).
This leads again to a condition that can always be satisfied for $a$
sufficiently small, whatever the dimension $d$.
Let $\xi_\star$ be the value of $\xi_{\text{min}}$ at which the existence
condition Eq.~(\ref{cond}) of
the $B$ phase gets violated when $\rho_0\rightarrow\rho_{0c}^+$. One readily
finds 
\begin{equation}\label{defxietoile}
\xi_\star=\Big(\frac{D_A}{k\rho_c^{1/2}}\Big)^{\tfrac{2}{4-d}}
\ln^{-\tfrac{2}{4-d}}\frac{D_A}{D_B}
\end{equation}
This is the spatial scale at which the instability sets in that causes the first
order transition. We also have to check that $\rho_{0c}\ll \rho_c$, in order to
be consistent with $\rho_0\ll \rho_c$ which was assumed following
Eq.~(\ref{defrho0}). This condition is certainly satisfied in the limit
$D_B\rightarrow D_A^-$, that we shall consider now. Setting as before
\begin{equation}
\sigma=1-\frac{D_B}{D_A}
\end{equation}
we obtain from the preceding equations
\begin{equation}\label{jumpsigma}
\rho_{0c}\simeq \text{cst.}\,(4-d)\,\sigma^{\tfrac{2}{4-d}} \qquad 
(\sigma\rightarrow 0^+)
\end{equation}
\begin{equation}\label{xistarsigma}
\xi_\star=\Big(\frac{D_A}{k\rho_c^{1/2}}\Big)^{\tfrac{2}{4-d}}
\,\sigma^{-\tfrac{2}{4-d}} \qquad (\sigma\rightarrow 0^+)
\end{equation}
The relaxation time towards zero of the $B$ density 
in the vicinity of the new critical density is
\begin{eqnarray}
T_B&\equiv&T_{B,\xi_\star}\nonumber\\
&\sim&\sigma^{-\tfrac{4}{4-d}} \qquad (\sigma\rightarrow 0^+)
\end{eqnarray}
Comparison of Eq.~(\ref{jumpsigma}) with the tricritical scaling predictions of 
the previous section
leads, with $\eps=4-d$, to the identification
\begin{equation}
\delta = \frac{\eps}{2}
\end{equation}
We expect that the exact theory gives power laws for the same quantities as the
heuristic theory does, although with different exponents. One reason for this is 
the difficulty of correctly keeping track of
the lattice parameter $a$.

\subsection{A nucleation picture}
Finally we would like to draw a 
parallel between the heuristic arguments developed above and
the kinetics of first order transitions~\cite{LandauX} in thermodynamic
systems. In those systems
the description
is based on a nucleation picture : the transition from a metastable to a stable
phase occurs as the result of fluctuations in a homogeneous medium. These
fluctuations permit the formation of
small quantities of a new phase, called nuclei. However the creation of an 
interface
is an energetically unfavored process, so that below a certain size nuclei
shrink and disappear. Nuclei having a size greater than a critical radius
$\xi_\star$ will survive an eventually expand. 
The analysis of the competition between bulk free energy and surface tension
leads to  an
estimate of a critical nucleus size.

In the original reaction-diffusion problem
there is of course no such concept as a bulk free 
energy or surface tension. Nuclei are analogous to regions that are free
of $B$ particles. Those analogies should not be
overinterpreted: they merely reinforce the intuitive picture of the
reaction-diffusion processes leading to the first-order transition.

\section{Simulations in two dimensions: a hysteresis loop}

In this section we present the results of simulations performed on a
two-dimensional 500$\times$500 lattice with periodic boundary conditions.
At the beginning of the simulation
particles are placed randomly and independently on the sites
of the lattice, with an average density $\rho=0.2$. The ratio of the $B$
particle density to the total density is
arbitrarily chosen equal to 0.3. The decay probability of the $B$ particles is
$\gamma=0.1$ per time step, and the contamination probability is $k=0.5$
per time step. These parameters are held
fixed. The diffusion constants $D_A$ and $D_B$ are varied.\\

In each time step the reaction-diffusion rules are implemented by the
following three operations. 
\begin{enumerate}
\item Each $B$ particle is turned into an $A$ with probability $\gamma$. 
\item Each $A$\, ($B$) particle moves with a probability $4D_A$\,
($4D_B$); a moving particle goes 
to a randomly chosen nearest neighbor site.
\item An $A$ particle is contaminated with probability $k$ by each of the $B$
particles on the same site. 
\end{enumerate}
Then the new value of the average $B$ density is evaluated and a new
time step is begun.
The process stops either when the system has fallen into its absorbing
state or when the $B$ density appears to have stabilized {\it "(active
state)"}. 
The latter situation is 
considered to be reached when the slope of $\rho_B(t)$, as
measured from a linear fit to the last 100 time steps, is $10^{-5}$ times as small as
the maximal variation of the density of those 100 points.\\

After this fixed density run we construct as follows a starting
configuration for a new run in which 
the total density is increased by a factor 1.004. Two situations may
occur. {\it If} at the end of the
run just terminated the $B$ particle density is {\it nonzero}, then we
obtain the new starting configuration from the final one of the preceding run
by randomly placing extra particles on
the lattice while keeping the ratio of $B$'s to the total number of
particles constant.
{\it If} at the end of the run just terminated the $B$ particle density
is {\it zero}, then we construct a new starting configuration with
a $B$ density equal to its value in the starting configuration of 
the preceding run.

A new run is carried out at the new density. This process
is iterated until some upper value of the total density is reached,
taken equal to $\rho=0.5$  in the
present simulations. After that we carried out a step-by-step decrease of the total
density, using a reduction factor of 0.996
per step, until we reached again the
total density $\rho=0.2$ of the beginning of the simulation.\\

\begin{figure}[h]
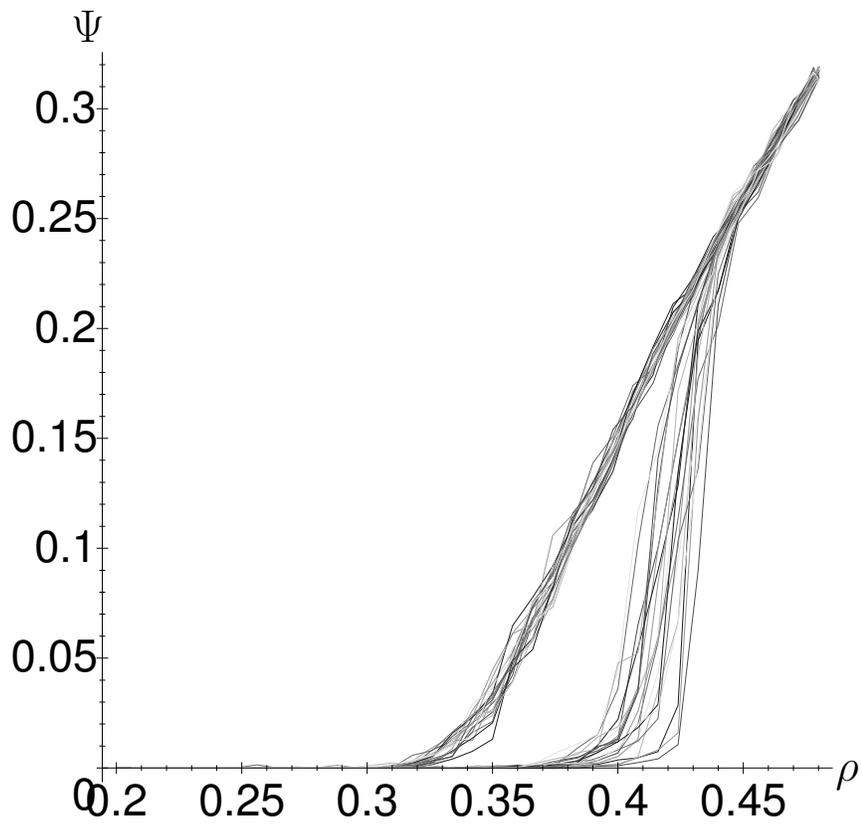

\caption{The order parameter $\Psi$ for $4D_A=0.8,$\,and\,$
4D_B=0.1$,\, with $\gamma
=0.1$ and $k=0.5$. The system cycles anticlockwise through a hysteresis loop.}
\label{leroy1}
\end{figure}

\begin{figure}[h]
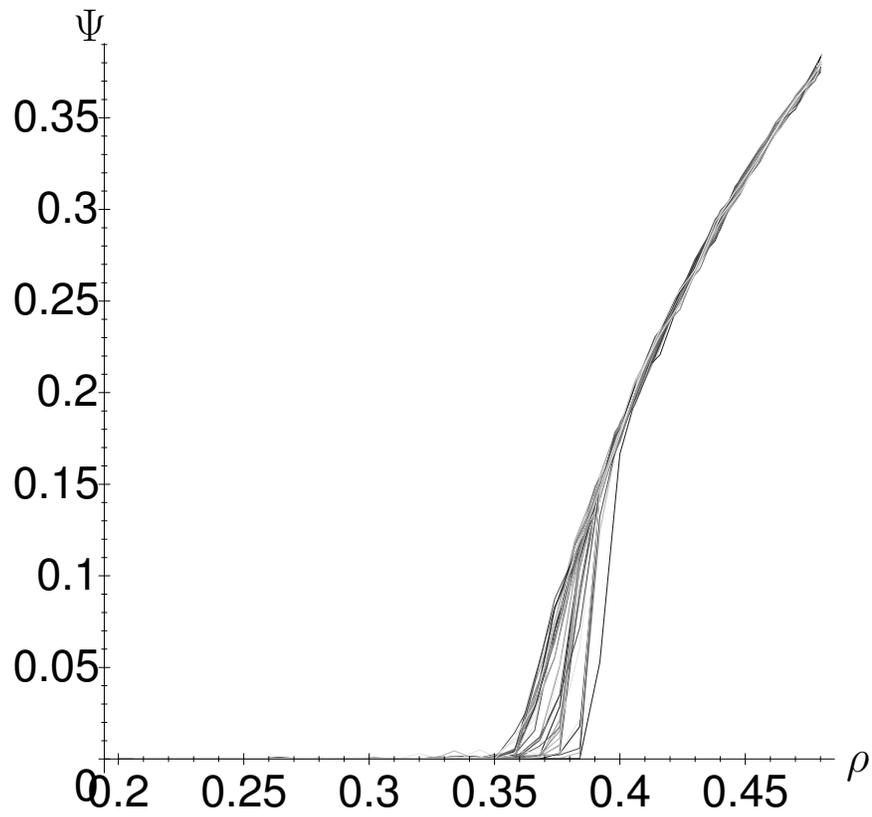

\caption{Same as Fig.\,2, but for
$4D_A=0.1,$\,and\,$4D_B=0.8$. In this case the transition is known to be
continuous and no hysteresis loop appears.}
\label{leroy2}
\end{figure}

This whole procedure constitutes a simulation cycle. In
this way we produced 21 cycles with different pseudo-random numbers. 
Figures \ref{leroy1} and \ref{leroy2} show the resulting order parameter
curves in two different cases: in Figure \ref{leroy1} we have $D_B<D_A$
and the system cycles counterclockwise through
a hysteresis loop, signalling the occurrence of a first
order transition. In Figure \ref{leroy2} exactly the same 
simulation procedure does lead to
some dispersion in the order parameter curves, but not to a clearcut
hysteresis loop; in
this case the transition is known to be continuous.

\section{Perturbative calculation of the equation of state}
\subsection{One-loop perturbation expansion}
In this section we determine the one-loop equation of state. We start from the
dynamic functional of Eq.~(\ref{actioncomplete}), in which we have included a 
particle source term
$-\int\dd^dr\dd t\;\lambda h(\br,t)\bar{\psi}(\br,t)$. The equations of motion 
for the fields are
\begin{eqnarray}
0 = \frac{\delta {S}}{\delta \bar{m}} =
\partial_{t} m - \lambda (\rho \nabla^{2} m + \sigma \nabla^{2}
\psi) + 2 \lambda \rho \nabla^{2} \bar{m} \label{eqmo1} \\
0 = \frac{\delta {S}}{\delta \bar{\psi}} =
\partial_{t} \psi + \lambda (\tau - \nabla^{2} + f m) \psi +
\frac{\lambda g}{2} \psi^{2} - \lambda \tilde{g} \psi \bar{\psi}
- \lambda h \label{eqmo2}
\end{eqnarray}
Note that the source term is not necessarily constant.
Equations~(\ref{eqmo1}) and~(\ref{eqmo2}) are valid when they
are inserted in averages. Taking the averages of (\ref{eqmo1})
and~(\ref{eqmo2}) we find for the densities
$M({\br} ) = \langle m({\br} , t) \rangle $ and $\Psi({\br} ) =
\langle\psi({\br} , t)\rangle$  in a stationary state the exact
equations
\begin{eqnarray}
\rho \nabla^{2} M(\br ) + \sigma \nabla^{2} \Psi(\br ) = 0
\label{eqM} \\
\left[ \tau - \nabla^{2} + f M(\br ) + \frac{g}{2} \Psi(\br )
\right] \Psi(\br ) + f C_{m \psi}(\br ) + \frac{g}{2}
C_{\psi}(\br ) = h(\br ) \label{eqPsi}
\end{eqnarray}
with the correlation functions
\begin{equation}
C_{m \psi}(\br ) = \left\langle \left(m-M(\br )\right)
\left(\psi-\Psi(\br )\right) \right\rangle \qquad
C_{\psi}(\br ) = \left\langle \left(\psi-\Psi(\br )\right)^{2}
\right\rangle . \label{corr}
\end{equation}
>From equation~(\ref{eqM}) it follows that
\begin{equation}
M(\br ) = -\frac{\sigma}{\rho} \Psi(\br ) + c(\br )
\end{equation}
where $c(\br )$ is a harmonic function ($\nabla^{2} c = 0$).
Here we assume that $c$ is constant. (It has to be constant in the
thermodynamic limit if both $M$ and $\Psi$ are free of singularities
and finite for $r \to \infty$.) If $\int_{V} \dd^{d}r\,m(\br,t) = 0$
(which can always be achieved by a shift of $\tau$) we get
\begin{equation}
c = \frac{\sigma}{\rho V} \int_{V} \dd^{d}r \Psi(\br ) 
\end{equation}
where $V$ denotes the volume of the system.

The mean field equation for the profile reads
\begin{equation}
\left[ \tau + f c - \nabla^{2} + \frac{\bar{g}}{2}
\Psi_{mf}(\br ) \right] \Psi_{mf}(\br ) = h(\br )
\label{mfprof}
\end{equation}
where $\bar{g} = g -
2\sigma f/\rho$. Equation~(\ref{mfprof}) shows that stability
of the mean field theory requires that $\bar{g} \geq 0$.
For negative ${\bar{g}}$ higher powers in $\psi$
have to be taken into account in the functional ${S}$. 
We first consider an external field $h$ which is constant within
a sphere of radius $R$ and vanishes for $r > R$. For simplicity
we take the thermodynamic limit $V, R \to \infty$ in such a way
that $R^{d}/V \to 0$. In this case the region outside the sphere
acts as a reservoir for the homogeneous mode of the field $m$,
and in~(\ref{mfprof}) we can set $c=0$ for $\tau > 0$.

To calculate the one-loop correction to the equation of state
we shift the fields $\psi$ and $m$ by their average values and
obtain
\begin{equation}
  {S}[\bar{m}, M+m; \bar{\psi}, \psi + \Psi] =
{S}_{0}[\bar{\psi}; \Psi] + {S}_{G}[\bar{m},
m; \bar{\psi}, \psi; \Psi] + {S}_{I}[m; \bar{\psi}, \psi]
\end{equation}
with
\begin{equation}
{S}_{0}[\bar{\psi}; \Psi] = \int \dd t \int \dd^{d}r \lambda
\bar{\psi} \left[ \left( \tau -\nabla^{2} + \frac{\bar{g}}{2}
\Psi \right) \Psi - q \right]
\end{equation}
where we have kept the full $\br $-dependence of $\Psi$ and
expressed $M$ in terms of $\Psi$.
The Gaussian part of ${S}$ reads
\begin{eqnarray}
  {S}_{G}[\bar{m}, m; \bar{\psi}, \psi; \Psi] = 
\int \dd t \int \dd^{d}r \left[
\bar{m} \left(\partial_{t} m - \lambda \nabla^{2} (\rho m +
\sigma \psi) \right) - \lambda \rho (\nabla \bar{m})^{2} \right.
\nonumber \\
\left. \bar{\psi} \left( \partial_{t} \psi + \lambda \left( \tau -
\nabla^{2} + ( g - \sigma f/\rho ) \Psi \right)
\psi + \lambda f \Psi m \right) - \frac{1}{2} \lambda \tilde{g}
\Psi \bar{\psi}^{2} \right] \label{JG}
\end{eqnarray}
and the interaction part is
\begin{equation}
{S}_{I}[m; \bar{\psi}, \psi] = \int \dd t \int \dd^{d}r \lambda
\left[ f m \psi \bar{\psi} + \frac{1}{2} \psi (g \psi - \tilde{g}
\bar{\psi}) \bar{\psi} \right].
\end{equation}
The Gaussian propagators $G_{\psi} = \langle \psi \bar{\psi}
\rangle$, $G_{m} = \langle m \bar{m} \rangle$, $G_{m \psi} =
\langle m \bar{\psi} \rangle$ and $G_{\psi m} = \langle \psi
\bar{m} \rangle$ follow from~(\ref{JG}). They satisfy the
differential equations
\begin{eqnarray}
  \left( \partial_{t} + \lambda (\bar{\tau} - \nabla^{2}) \right)
G_{\psi}(\br , \br ^{\prime}; t-t^{\prime}) + \lambda f \Psi
G_{m\psi}(\br , \br ^{\prime}; t-t^{\prime})
= \delta^{(d)}( \br  - \br ^{\prime}) \delta(t-t^{\prime})
\label {Gpsi} \\
  \left( \partial_{t} - \lambda \rho \nabla^{2} \right)
G_{m\psi}(\br , \br ^{\prime}; t-t^{\prime}) -
\lambda \sigma \nabla^{2} G_{\psi}(\br , \br ^{\prime};
t-t^{\prime}) = 0 \\
  \left( \partial_{t} - \lambda \rho \nabla^{2} \right)
G_{m}(\br , \br ^{\prime}; t-t^{\prime}) -
\lambda \sigma \nabla^{2} G_{\psi m}(\br , \br ^{\prime};
t-t^{\prime}) = \delta^{(d)}( \br  - \br ^{\prime})
\delta(t-t^{\prime}) \\
\left( \partial_{t} + \lambda (\bar{\tau} - \nabla^{2}) \right)
G_{\psi m}(\br , \br ^{\prime}; t-t^{\prime}) +
\lambda f \Psi G_{m}(\br , \br ^{\prime}; t-t^{\prime}) = 0
\label{Gpsim}
\end{eqnarray}
with $\bar{\tau} = \tau + (g - \sigma f/\rho) \Psi$.
The Gaussian propagators can be used to determine the equal time correlation
functions~(\ref{corr}) to lowest nontrivial order:
\begin{eqnarray}
  C_{m\psi}(\br ) = \int_{0}^{\infty} \dd t \int
\dd^{d}r^{\prime} \left[ \lambda \tilde{g} \Psi(\br ^{\prime})
G_{m \psi}(\br , \br ^{\prime}; t) G_{\psi}(\br ,
\br ^{\prime}; t) \right. \nonumber \\
\left. + 2 \lambda \rho \left(\nabla^{\prime} G_{m}(\br ,
\br ^{\prime}; t) \right) \left(\nabla^{\prime} G_{\psi
m}(\br , \br ^{\prime}; t) \right)\right] \\
  C_{\psi}(\br ) = \int_{0}^{\infty} \dd t \int
\dd^{d}r^{\prime} \left[ \lambda \tilde{g} \Psi(\br ^{\prime})
\left( G_{\psi}(\br , \br ^{\prime}; t) \right)^{2}
+ 2 \lambda \rho \left( \nabla^{\prime} G_{\psi m}(\br ,
\br ^{\prime}; t)\right)^{2} \right] .
\end{eqnarray}
For constant $\Psi$ the equations~(\ref{Gpsi})-(\ref{Gpsim}) can easily
be solved by Fourier transformation. In this way one obtains
for the fluctuation term in the equation of state~(\ref{eqPsi})
\begin{equation}
  f C_{m\psi} + \frac{g}{2} C_{\psi} = \int_{\bf q}
\frac{\Psi}{\bar{\tau} + (1+\rho) q^2} \left[ \frac{1}{4} \tilde{g}
g - f^{2} + \frac{\bar{g} (\tilde{g} \rho q^{2} + 2 f^{2} \Psi)}{4
(\tau^{\prime} + q^2)} \right]
\end{equation}
with $\tau^{\prime} = \bar{\tau} - \sigma f \Psi/\rho$.
After dimensional regularization the momentum integral becomes
\begin{eqnarray}
  f C_{m\psi} + \frac{g}{2} C_{\psi} = -
\frac{A_{\eps} \Psi}{2 \eps (1+\rho)} \left(
\frac{\bar{\tau}}{1+\rho}\right)^{1-\eps/2} \left[
\tilde{g}g - 4 f^{2} + \tilde{g}\bar{g}\rho 
+ \frac{1+\rho}{\bar{\tau}} \bar{g} (\tilde{g}
\rho \tau^{\prime} - 2 f^{2} \Psi ) \right. \nonumber \\
\left. + \frac{\eps}{2} \,\frac{(1+\rho)^{2} \tau^{\prime}}{
\bar{\tau}} \, \frac{\bar{g} (\tilde{g}
\rho \tau^{\prime} - 2 f^{2} \Psi )}{\bar{\tau} - (1+\rho)
\tau^{\prime}} \ln \frac{(1+\rho)\tau^{\prime}}{\bar{\tau}}
+ {\cal O}(\eps^{2}) \right] \label{Cregul}
\end{eqnarray}
where $A_{\eps} = (4\pi)^{-d/2}\Gamma(1+\eps/2)/(1 -
\eps/2)$.
\subsection{Renormalized equation of state}
In order to absorb the $\eps$-poles in the equation of
state~(\ref{eqPsi},\ref{Cregul}) into a reparametrization
of coupling constants and fields we make use of the renormalized
quantities $\psi_{\sR}$, $\lambda_{\sR}$, $\rho_{\sR}$ etc. introduced in
Eqs.~(\ref{Rcouplings},\ref{Zfactors}).

The renormalized quantities satisfy the equation of state
\begin{eqnarray}\label{equationdetat}
  h_{\sR} = \Psi_{\sR} \left\{ \tau_{\sR}+\frac{\bar{g}_{\sR}}{2}
\Psi_{\sR}+\frac{1}{4 (1+\rho_{\sR})} \left[ \left((1+\rho_{\sR})u-4v-
2w\right) \frac{\bar{\tau}_{\sR}}{1+\rho_{\sR}} \ln \frac{\mu^{-2}
\bar{\tau}_{\sR}}{1+\rho_{\sR}} \right. \right. \label{one-loop} \\
\left. \left. + \frac{(\rho_{\sR} u - 2w)\tau_{\sR}^{\prime} -
2 v \bar{g}_{\sR}\Psi_{\sR}}{\bar{\tau}_{\sR}-(1+\rho_{\sR})
\tau^{\prime}_{\sR}} \left( \bar{\tau}_{\sR} \ln \frac{\mu^{-2}
\bar{\tau}_{\sR}}{1+\rho_{\sR}} - (1+\rho_{\sR}) \tau_{\sR}^{\prime}
\ln (\mu^{-2} \tau_{\sR}^{\prime}) \right) \right] \right\} \nonumber
\end{eqnarray}
where $\bar{g}_{\sR} = g_{\sR} - 2f_{\sR}/\rho_{\sR}$, 
$\bar{\tau}_{\sR}=\tau_{\sR} + (g_{\sR}-f_{\sR}/\rho_{\sR}) \Psi_{\sR}$ and
$\tau_{\sR}^{\prime} = \tau_{\sR} + \bar{g}_{\sR}\Psi_{\sR}$ are the
renormalized counterparts of $\bar{g}$, $\bar{\tau}$ and
$\tau^{\prime}$, respectively.
For $f_{\sR}=v=w=0$ we recover the one  loop-equation of state for
directed percolation~\cite{JKO}. To simplify the writing in
equation~(\ref{one-loop}) the geometrical factor $A_{\eps}$,
the momentum scale $\mu$ and $\sigma_{\sR}$ have been absorbed into a
rescaling of $\Psi_{\sR}$ and $h_{\sR}$, i.e.
\begin{equation}
A_{\eps}^{-1/2} \mu^{\eps/2} \sigma_{\sR} \Psi_{\sR} \to
\Psi_{\sR} \text{ and }
A_{\eps}^{-1/2} \mu^{\eps/2} \sigma_{\sR} h_{\sR} \to
h_{\sR} 
\end{equation}
Hereafter we will drop the index ``R'' since only renormalized
quantities will be used.

\section{Flow equations} \label{flowequations}
\subsection{Renormalization group for the equation of state}
The renormalizability of the field theory implies a set partial
differential equations for the vertex functions.  These are the
renormalization group equations which follow from the
independence of the bare vertex functions on the momentum scale
$\mu$. To investigate the equation of state in the critical region
we need the one-point vertex function $\Gamma^{(1,0)}$ which is (up to a
factor $\lambda$) equal to $h(\tau, \Psi; u, v, w, \rho; \mu)$. The 
renormalization group equation for $h$ reads
\begin{equation}
  \left[\mu \frac{\partial}{\partial\mu} + \beta_{u}
\frac{\partial}{\partial u} + \beta_{v} \frac{\partial}{\partial v}
+ \beta_{w} \frac{\partial}{\partial w} + \beta_{\rho} 
\frac{\partial}{\partial \rho} + \kappa \tau \frac{\partial}{\partial
\tau} + \zeta - \gamma \right] h(\tau, \Psi; u, v, w, \rho; \mu) = 0
\label{RGE} 
\end{equation}
with the $\beta$-functions given in Sec.~2.
\subsection{Scaling form of the equation of state}
\begin{figure}[h]
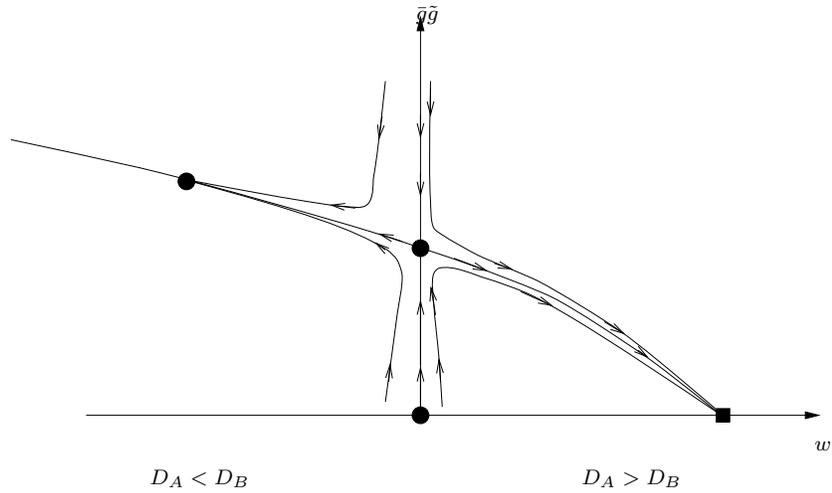

\caption{Flow diagram in the $(\bar{g}\tilde{g},w)$ plane. The leftmost black
dot stands for the $D_A<D_B$ fixed point while the one lying on the $w=0$ axis
stands for the symmetric $D_A=D_B$ fixed point. They both describe second order
transitions. Typical trajectories have been drawn. Those starting close to the
symmetric fixed point but with an initial positive $w$ eventually flow away from 
the
symmetric fixed point.}
\label{figure3}
\end{figure}

The renormalization group equation Eq.~(\ref{RGE}) can be solved by 
characteristics with the
result
\begin{eqnarray}
  h(\tau, \Psi; u, v, w, \rho; \mu) =
Y_{h}(\ell)^{-1} (\mu \ell)^{2+d/2} \nonumber \\
\times h\left(Y_{\tau}(\ell) (\mu \ell)^{-2} \tau,
(\mu \ell)^{-d/2} \Psi; u(\ell), v(\ell), w(\ell), \rho(\ell); 1\right) 
\label{flow}
\end{eqnarray}
where
\begin{eqnarray}
\frac{\dd a(l)}{\dd \ln \ell} = \beta_{a}(u(\ell), v(\ell), w(\ell), \rho(\ell))
\qquad \mbox{for $a = u, v, w, \rho$} \label{fleqs} \\
\frac{\dd \ln Y_{\tau}(\ell)}{\dd \ln \ell} = \kappa(u(\ell), v(\ell), w(\ell),
\rho(\ell)) \\
\frac{\dd \ln Y_{h}(\ell)}{\dd \ln \ell} = \gamma(u(\ell), v(\ell), w(\ell), 
\rho(\ell))-
\zeta(u(\ell), v(\ell), w(\ell), \rho(\ell))
\end{eqnarray}
are the characteristics with the initial conditions $u(1)=u$,
$v(1)=v$, $w(1)=w$, $\rho(1) = \rho$ and $Y_{\tau}(1)=Y_{h}(1)=1$.
Some solutions of the flow equations~(\ref{fleqs}) are depicted in
figure~\ref{figure3}. For small initial values of $w \propto D_{A}-
D_{B}$ the trajectories first approach the unstable manifold of the
symmetric fixed point ($w=0$) before they flow away. The unstable manifold
leaves the stability region of the mean field theory ($\bar{g}>0$, or
$u-2 w/\rho > 0$) at the point $(u_{\star}, v_{\star}, w_{\star},
\rho_{\star})$. The numerical solution of the flow equations
yields $u_{\star} = 12.32 \eps$, $v_{\star}=2.104\eps$,
$w_{\star} = 22.11 \eps$, and $\rho_{\star} = 3.589$.
The intersection point of the unstable manifold with the stability
edge $\bar{g}=0$ is of special interest since the perturbatively
improved mean field theory should be a good approximation for
small $\bar{g}$ (as will be discussed in the next section). Therefore
the phase transition for very small $w>0$ is governed by the
equation of state at the point $(u_{\star}, v_{\star}, w_{\star},
\rho_{\star})$. In the following we shall denote by $\ell_\star$ the value of
the flow parameter at which $u(\ell_\star)-2w(\ell_\star)/\rho(\ell_\star)=0$
and by $\xi_\star=\ee^{\ell_\star}$ the related length scale. One can identify
$\xi_\star$ with its heuristic counterpart defined in Eq.~(\ref{defxietoile}).
The flow equations are too complicated to be solved analytically for all $\ell$,
however it is possible, using scaling arguments, to predict that, as
$\sigma\rightarrow 0^+$, $\xi_\star\propto \sigma^{2/\gamma_s}$, a form also
proposed in Eq.~(\ref{xistarsigma}).  


\section{A first-order transition for $D_{A} > D_{B}$}

In this section we study the mean field equation of state with the
one-loop fluctuation correction~(\ref{one-loop}) for small values
of the coupling $\bar{g} \ll g$. Our motivation is an analogy
of our reaction diffusion system with spin systems with cubic
anisotropy. Using a modified Ginzburg criterion for systems with
cubic anisotropy Rudnick~\cite{Rudnick} (see also~\cite{AY}) has
shown that the fluctuation corrected mean field approximation should
give reliable results if the values of the coupling coefficients are
close to the stability edge of the mean field theory. In the case of
the reaction diffusion system with $D_{A} > D_{B}$ the stability edge
is given by $\bar{g}=0$.


Now assume that we start with a very small value of the coupling $w$
and choose the flow parameter $\ell=\ell_\star$ in~(\ref{flow}). After the
$\ell$-dependent prefactors $Y_{\tau}(\ell_\star)
\ell_\star^{-2}$ etc. have been asborbed into a rescaling of $\tau$, $h$, and
$\Psi$ the improved mean field equation of state takes the
simple form
\begin{equation}
h = \Psi \left[ \tau + \frac{u_{\star}-4 v_{\star}}{4 (1+
\rho_{\star})^{2}} \left( \tau + \frac{g_{\star}}{2} \Psi \right)
\ln \frac{\tau + g_{\star} \Psi/2}{\mu^{2}(1+\rho_{\star})} +
{\cal O}(\mbox{two-loop}) \right] \label{1st-ord}
\end{equation}
with $(u_{\star}-4 v_{\star})/(4 (1+ \rho_{\star})^{2}) = 0.04635
\,\eps > 0$.

\begin{figure}[h]
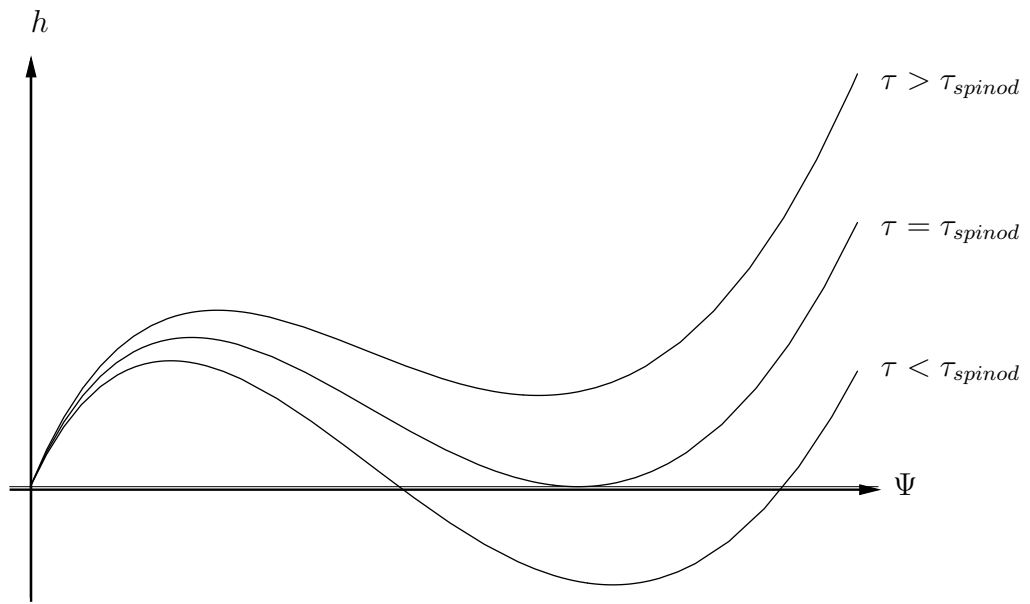

\caption{Sketch of the function $h(\tau, \Psi)$,
Eq.~(\ref{1st-ord}).}
\label{sketch}
\end{figure}

In the limit $h \to 0^+$ the absorbing state with $\Psi = 0$ is a
solution of the equation of state for
all $\tau$. For $\tau < \tau_{spinod}$ with
\begin{equation}
\tau_{spinod} = \mu^{2} \ee^{-1} \frac{u_{\star}-4v_{\star}}{4
(1+\rho_{\star})} = {\cal O}(\eps)
\end{equation}
there is also a solution with $\Psi > 0$ and $\partial \Psi / \partial h
> 0$ (see figure~\ref{sketch}). To see this one should anticipate that when the
order parameter is of the order of its value at $\tau_{spinod}$  one has
\begin{equation}
\frac{\tau_{spinod}}{g_\star\Psi}={\cal O}(\eps)
\end{equation}
and to leading order the reasoning is carried out on
\begin{equation}\label{eqetatsimple}
h=\Psi\left[\tau+\frac{u_{\star}-4v_\star}{4(1+\rho_\star)^2}\frac{g_\star}{2}
\Psi\ln\frac{g_\star\Psi}{2(1+\rho_\star)\mu^2}\right]
\end{equation}
Solving the latter equation for $h=0$ and $\frac{\p h}{\p \Psi}=0$ yields, to
leading order in $\eps$,
\begin{equation}
\Psi_{spinod}=\frac{8(1+\rho_\star)^2}{u_\star-4v_\star}\frac{\tau_{spinod}}
{g_\star},\;\;\;\tau_{spinod}=\mu^{2} \ee^{-1} \frac{u_{\star}-4v_{\star}}{4
(1+\rho_{\star})}
\end{equation}
which {\it a posteriori} justifies working on the approximation
Eq.~(\ref{eqetatsimple}). Since the susceptibility $\chi=\partial \Psi /
\partial h$ (the response of the order parameter to particule injection) is 
positive
the new solution is at least metastable for all $\tau < \tau_{spinod}$.
The open question is for which range of $\tau$ the solution is
stable, i.e., stable also with respect to nucleation processes.
If we could derive the equation of state from a free energy this
question would be easy to answer: a metastable solution is a {\em local}
minimum of the free energy while a stable solution is a {\em global}
minimum. However, since there is no free energy for the reaction diffusion
system we have to look for a different stability criterion, e.g. the form
of density profiles. By analogy with equilibrium systems we expect that
the solution with $\Psi>0$ is stable below a coexistence ``temperature" (in our
case we should talk about a coexistence density)
$\tau_{coex}$ with $0 < \tau_{coex} < \tau_{spinod}$.

\section{Deriving $\tau_{coex}$ from a density profile}

If $\tau$ is so large that equation~(\ref{1st-ord}) has only the
trivial solution $\Psi = 0$ the density $\Psi(\br )$ generated by
a local source $h(\br )$ decays rapidly with increasing distance
from the region in which $h(\br )$ is nonzero. Consider for
instance a plane particle source with
\begin{equation}
h(\br ) = h_{0} \delta(r_{\bot}) \label{source}
\end{equation}
where $r_{\bot}$ is the coordinate perpendicular to the source.
For simplicity assume that $h_{0} \to \infty$. For large $r_{\bot}$
the profile [i.e., the solution of~(\ref{eqPsi})] will become
independent of $r_{\bot}$ and tend to either (a) $\Psi = 0$ if
$\tau$ is sufficiently large or (b) the nonzero solution
of~(\ref{1st-ord}). In the thermodynamic limit the whole profile
$\Psi(\br )$ is uniquely determined by the source $h(\br )$
since $\Psi$  has to be finite for $r_{\bot} \to \pm\infty$.
In the case (a) we can conclude that $\tau > \tau_{coex}$
whereas (b) occurs  for $\tau < \tau_{coex}$.

In order to calculate a density profile perturbatively one usually
starts with the mean field profile and assumes that the fluctuation
corrections are small (of the order $\eps$, say). In the present
case this procedure would not lead to the desired result since the
nonzero solution of the equation of state for $\tau < \tau_{spinod}$
is of the order $\eps^{0}$. It can therefore not be derived
as a small correction to $\Psi_{mf}$. Instead we have to compute the
{\em equation for the density profile} perturbatively and then study
the asymptotic behavior of its solutions. This means that we need
the correlation functions $C_{m\psi}(\br ; \{\Psi\})$ and
$C_{\psi}(\br ; \{\Psi\})$ in equation~(\ref{eqPsi}) for a general
function $\Psi(\br )$. Of course, we cannot compute these
functions exactly but it is possible to derive a systematic gradient
expansion for $C_{m\psi}$ and $C_{\psi}$.

We first compare the leading $\eps$-orders in equation~(\ref{eqPsi})
(with $M = -\sigma \Psi/\rho$): Since we expect that
the coexistence point to be located in the parameter range $0 < \tau \leq
\tau_{spinod} = {\cal O}(\eps)$ we may set $\tau = {\cal O}(\eps)$.
The limit of a very weak first order transition is governed by
the point $(u_{\star}, v_{\star}, w_{\star}, \rho_{\star})$ which
means that $\bar{g}=0$ and that the combination $f C_{m\psi}+ \frac{g}{2} 
C_{\psi} $ is of
order ${\cal O}({\eps})$.
Comparing the terms on the l.h.s. of~(\ref{eqPsi}) therefore yields
$(\nabla^{2} \Psi)/\Psi = {\cal O}({\eps})$, i.e. gradients of $\Psi$
may be considered as small quantities when we calculate $C_{m\psi}$ and
$C_{\psi}$. Using the Taylor series
\begin{equation}
\Psi(\br ^{\prime}) = \Psi(\br ) + \sum_{N=1}^{\infty}
\frac{1}{N!} \sum_{\alpha_{1} \cdots \alpha_{N}} (r^{\prime}-
r)_{\alpha_{1}} \ldots (r^{\prime}-r)_{\alpha_{N}}
\partial_{\alpha_{1}} \ldots \partial_{\alpha_{N}} \Psi(\br )
\end{equation}
for the profile one arrives at a gradient expansion for $C_{m\psi}$
and $C_{\psi}$ of the form
\begin{equation}
  C_{\psi}(\br , \{\Psi\}) = C_{\psi}(\Psi(\br )) +
\sum_{N=1}^{\infty} \frac{1}{N!} \sum_{\alpha_{1} \cdots \alpha_{N}}
C_{\psi; \alpha_{1} \cdots \alpha_{N}}(\Psi(\br ))
\partial_{\alpha_{1}} \ldots \partial_{\alpha_{N}} \Psi(\br ) .
\label{gradexp}
\end{equation}
At leading order in $\eps$ only the first term on the r.h.s. of~%
(\ref{gradexp}) contributes, i.e., we may simply replace $\Psi$
in~(\ref{Cregul}) by the profile $\Psi(\br )$ and use the result
in equation~(\ref{eqPsi}). After application of the renormalization
group as before and absorbing $\ell$-dependent prefactors the equation
for the profile becomes
\begin{equation}
  h(\br ) = \Psi(\br ) \left[ \tau + \frac{u_{\star}-4 v_{\star}
}{4(1+\rho_{\star})^{2}} \left( \tau + \frac{g_{\star}}{2} \Psi(\br )
\right) \ln \frac{\tau + g_{\star} \Psi(\br )/2}{\mu^{2}
(1+\rho_{\star})} \right] - \nabla^{2} \Psi(\br ) +
{\cal O}(\eps^{2}) . \label{1st-ord-prof}
\end{equation}
In order to extend this result to the next order in $\eps$
one has to (i) compute $C_{m\psi}$ and $C_{\psi}$ for constant $\Psi$
to two-loop order and (ii) take into account the $\nabla^{2}
\Psi$-correction in the gradient expansion~(\ref{gradexp}) to one-loop
order.  In this way the $\nabla^{2}\Psi$-term in~(\ref{1st-ord-prof})
may receive a $\Psi$-dependent correction.

\begin{figure}[h]
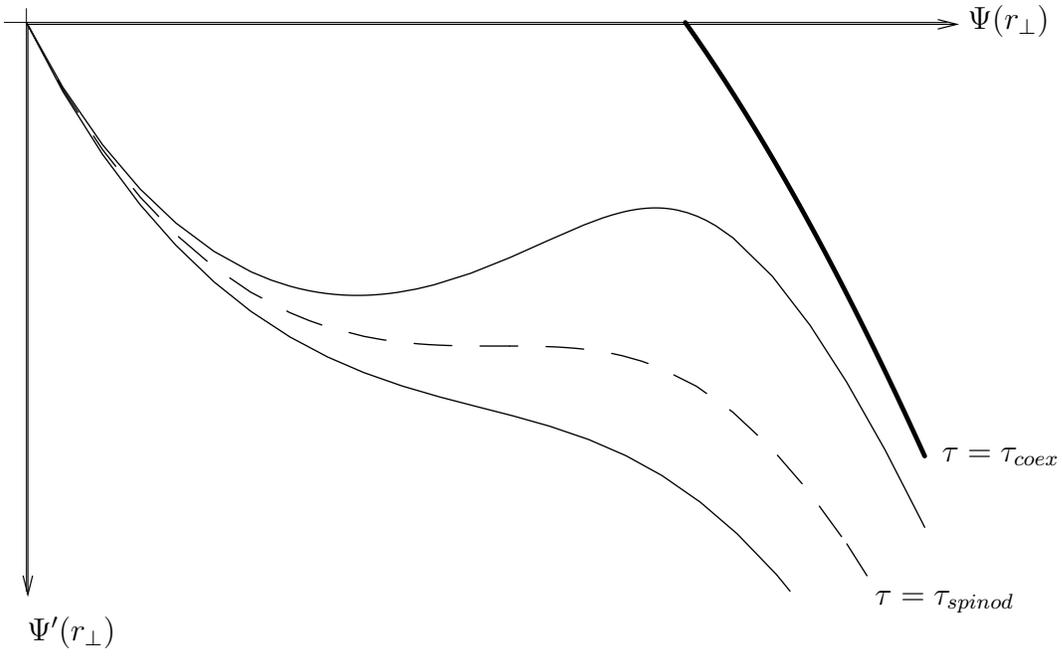

\caption{Derivative of the density profile $\Psi'(r_\perp)$ as a function of
$\Psi(r_\perp)$ for $\tau\geq
\tau_{coex}$. The boundary condition at $r_\perp=0^+$ is given by
$\Psi^{\prime}(0) = -h_{0}/2$.}
\label{sketch3}
\end{figure}

For $h(\br ) = h_{0} \delta(r_{\bot})$ Eq.~(\ref{1st-ord-prof})
can be integrated once after multiplication of both sides with
$\Psi^{\prime}(r_{\bot})$. In figure~\ref{sketch3} the result is depicted
for various values of $\tau$. There is a value $\tau_{coex} < \tau_{spinod}$
such that the profile does not tend to zero for $r_{\bot} \to \pm\infty$
if $\tau \leq \tau_{coex}$. As discussed above $\tau_{coex}$ is the
coexistence point below which the active phase becomes a stable solution
of the equation of state. Again one works with Eq.~(\ref{eqetatsimple}). In
practice one has to solve the system composed of Eq.~(\ref{eqetatsimple}) with 
$h=0$ along
with its integrated counterpart
\begin{equation}
0=\frac{\tau_{coex}}{2}\Psi_{coex}^2+
\frac{u_\star-4v_\star}{24(1+\rho_\star)^2}g_\star\Psi_{coex}^3\left[\ln\frac{g_
\star
\Psi_{coex}}{2\mu^2(1+\rho_\star)}-\frac
13\right]
\end{equation}
The explicit calculation yields 
$\Psi_{coex}=\frac{12(1+\rho_\star)^2\tau_{coex}}{(u_\star-4v_\star)g_\star}$ 
(which also
justifies the use of the simplified equation of state Eq.~(\ref{eqetatsimple})),
with
\begin{equation}
\tau_{coex} = \mu^{2} \ee^{-2/3} \frac{u_{\star}-4 v_{\star}}{6
(1+\rho_{\star})} = 0.93 \tau_{spinod} .
\end{equation}
$\tau_{coex}/\tau_{spinod}$ is not a universal number, but the susceptibility
ratio $\chi_{+}/\chi_{-}$ with
\begin{equation}
\chi_{+} = \lim_{h \to 0} \left.\frac{\partial \Psi}{\partial h}
\right|_{\Psi=0,\tau_{coex}} \qquad \chi_{-} = \lim_{h \to 0}
\left.\frac{\partial \Psi}{\partial h} \right|_{\Psi>0,\tau_{coex}} 
\end{equation}
is universal. To one-loop order one finds
\begin{equation}
\frac{\chi_{+}}{\chi_{-}} = \frac{1}{2} + {\cal O}(\eps)
\end{equation}
This result is analogous to the universality of the magnetic susceptibility
ratio found by Rudnick~\cite{Rudnick} and Arnold and Yaffe~\cite{AY}.
\section{Conclusions and prospects}
\subsection{A heuristic functional for the steady state phase diagram}
\begin{figure}[h]
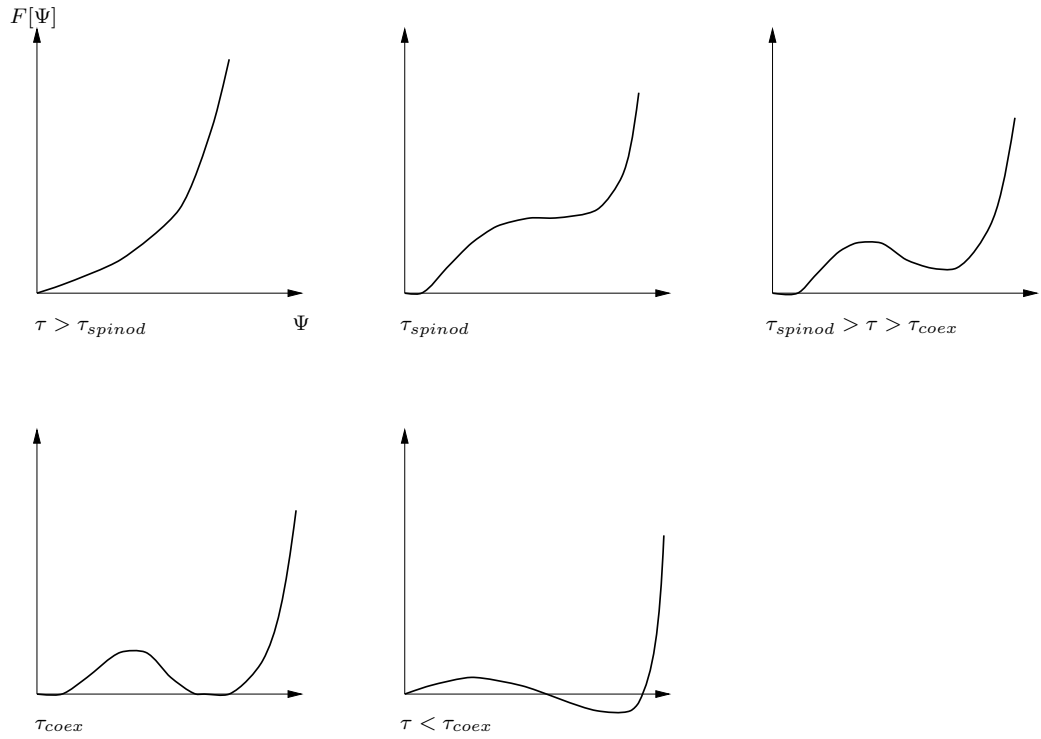
\caption{$F[\Psi]$ as a function of $\Psi$ for decreasing 
values of $\tau$. We
identify the spinodal point as the value of $\tau$ below which $F$ develops a
local nonzero minimum and the coexistence point as
the point at which the two minima become degenerate.}
\label{figure5}
\end{figure}

In this  paragraph we would like to build {\it a posteriori} a functional of the
order parameter field $\Psi(\br)$ describing the phase diagram in the stationary
state of the system. We emphasize that the following is only valid to one loop
order. We define $F[\Psi]$ by
\begin{equation}
F[\Psi]=\int\dd^dr\;\left[\frac{1}{2}(\nabla\Psi)^2+
\int_0^{\Psi({\br})}\!\!\!\!\!\!\!\!\dd\psi\,\,\Gamma^{(1,0)}[0,\psi]\right]
\end{equation}
By construction of course $\frac{\delta F}{\delta\Psi}=0$ is equivalent to the
equation of state Eq.~(\ref{eqetatsimple}). It is instructive to plot $F$ as a 
function of
$\Psi$ for various values of $\tau$ (see figure \ref{figure5}). There it appears
possible to deduce the phase diagram from the global minima of $F[\Psi]$. 
However the
route leading to the functional $F[\Psi]$ follows a series of field-theoretic
detours. The suggestive notation $F$, which reminds of a free energy (in the
equilibrium statistical mechanics sense) is however misleading. For instance it
could not be used as an effective Landau hamiltonian for the calculation of a
Gibbs partition function describing
fluctuations directly in the steady state. This functional is a remarkably 
compact
and intuitive way of summarizing the properties of the steady state phase 
diagram.
In particular the spinodal point $\tau_{spinod}$ appears as the point below 
which
$F$ develops a second minimum. Below $\tau_{coex}$ that minimum
becomes the global minimum. To one loop order the equilibrium vocabulary can 
therefore be used
carelessly.

\subsection{Summary}
In the course of this work we have elaborated a phenomenological description of
a fluctuation-induced first-order transition taking place in a nonequilibrium
steady state, the first one of this sort. We have in parallel applied 
field-theoretic
techniques to derive an effective (renormalization-group improved) equation of 
state that incorporates those
fluctuations. This yields a PDE for the order parameter in the steady state.
Performing a study of the stability (against space fluctuations of the order
parameter) of the solutions of this PDE has led us to a complete description of
the phase diagram. We have identified in this nonequilibrium situation a concept
analogous to the point of spinodal decomposition consistent with our
phenomenological description.

\subsection{Possible applications}
Among the many nonequilibrium systems that appear in the literature, driven 
diffusive
systems lend themselves to an analytic treatment by techniques 
similar to
those of the present article~\cite{BeateZia}. In a number of 
such
systems though, a significant portion of the phase space (in terms of control
parameters) escapes conventional analysis. In some cases we believe that the 
reason is
the occurrence of fluctuation-induced first order transition, such as in
\cite{BeateJanssen} or \cite{KornissZiaBeate}. It would be quite interesting to 
see how
both the technical argument and the heuristics can be extended to those systems. 
This
will be the subject of future work.\\\\\\

\noindent {\bf Acknowledgments: } K.~O. would like thank the 
Sonderforschungsbereich 237 of
the DFG for support and H.~K. Janssen for interesting discussions.

\newpage
\section*{Appendix}
There are two ways to proceed in order to obtain the equation of state to 
one-loop
order. In this appendix we follow the route familiar from static 
critical phenomena. 
We use the action Eq.~(\ref{Actioneffpourpsi}) as the starting point of our 
analysis. The first task is to determine
the one-loop expression of the effective potential $\Gamma$.
\begin{equation}
\Gamma[\bar{\psi},\psi]=S[\bar{\psi},\psi]+\frac 12\int\frac{\dd^d q}{(2\pi)^d}
\frac{\dd\omega}{2\pi}\ln\det{S''({\bq},\omega)}[\bar{\psi},\psi]
\end{equation}
with the matrix $S''$ defined by
\begin{equation}
{S}''({\bq},\omega)=\left(
\begin{array}{cc}
\frac{\delta^2 S}{\delta\bar{\psi}(-\bq,-\omega) \delta\psi(\bq,\omega)}&
\frac{\delta^2 S}{\delta\bar{\psi}(-\bq,-\omega) \delta\bar{\psi}(\bq,\omega)}
\\
\frac{\delta^2 S}{\delta{\psi}(-\bq,-\omega) \delta\psi(\bq,\omega)}&
\frac{\delta^2 S}{\delta{\psi}(-\bq,-\omega) \delta\bar{\psi}(\bq,\omega)}
\end{array}
\right)
\end{equation}
\begin{eqnarray}
{S}''_{11}({\bq},\omega)=
\lambda(\bq^2+\bar{\tau})-i\omega-\lambda\tilde{g}\Psi-\lambda^2\sigma
f\Psi\frac{\bq^2}{\bq^2-i\omega}-2\frac{(\lambda
f)^2\bq^2}{\bq^4+\omega^2}\Psi\bar{\psi}\\
{S}''_{12}({\bq},\omega)=-\frac{2(\lambda
f)^2\bq^2}{\bq^4+\omega^2}\Psi^2-\lambda\tilde{g}\Psi
\\
{S}''_{21}({\bq},\omega)=-\frac{2(\lambda
f)^2\bq^2}{\bq^4+\omega^2}{\bar{\psi}}^2-2\frac{\lambda^2\sigma f 
q^4}{q^4+\omega^2}\bar{\psi}
+\lambda g\bar{\psi}\\
{S}''_{22}({\bq},\omega)=\lambda(\bq^2+\bar{\tau})+i\omega-\lambda\tilde{g}\Psi-
\lambda^2\sigma
f\Psi\frac{\bq^2}{\bq^2+i\omega}-2\frac{(\lambda
f)^2q^2}{\bq^4+\omega^2}\Psi\bar{\psi}
\end{eqnarray}
For a homogeneous source term $h$ the equation of state for a homogeneous order 
parameter $\Psi$ now follows from the
requirement that 
\begin{equation}
\frac{\delta \Gamma}{\delta\bar{\psi}}[0,\Psi]=0
\end{equation}
It is a tedious but straighforward task to find the
one-loop correction to $\Gamma^{(1,0)}$ in the form of an integral over momentum
and frequency. Writing
\begin{equation}
\Gamma^{(1,0)}[\bar{\psi}=0,\psi=\Psi]=-\lambda h+\lambda
\Psi(\tau+\frac 12\bar{g}\Psi)+\delta\Gamma^{(1,0)}
\end{equation}
one finds
\begin{equation}\label{1loopGamma10}\begin{split}
\delta \Gamma^{(1,0)}=&-\lambda^2\int\frac{\dd^d q}{(2\pi)^d}
\frac{\dd\omega}{2\pi}
\left({\bq}^2+\bar{\tau}-\frac 12 g\Psi\right)\\
&\times
\left(\tilde{g}({\bq}^4+\omega^2)+2\lambda f^2{\bq}^2\Psi\right)\\
&
\times\left[\left(\omega^2-iA\omega-B\right)
\left(\omega^2+iA\omega-B\right)\right]^{-1}
\end{split}\end{equation}
where we have defined the auxiliary variables
\begin{equation}
A\equiv(\lambda+1){\bq}^2+\lambda\bar{\tau},\;\;\;B\equiv
{\bq}^2\lambda({\bq}^2+\tau')
\end{equation}
Upon using the following integration formulas,
\begin{equation}
\int\frac{\dd\omega}{2\pi}\frac{1}{\omega^2\pm i A\omega -B}=0,\;\;\;
\int\frac{\dd\omega}{2\pi}\frac{1}{|\omega^2\pm i A\omega -B|^2}=\frac{1}{2AB}
\end{equation}
Eq.~(\ref{1loopGamma10}) simplifies into
\begin{equation}\begin{split}
\delta \Gamma^{(1,0)}=&-\frac{\lambda}{2}\int\frac{\dd^d q}{(2\pi)^d}
\left[({\bq}^2+\bar{\tau}-\frac 12
g\Psi)\left(\tilde{g}({\bq}^2+\frac{\tau'}{1+\rho})+\frac{2f^2}{1+\rho}
\Psi\right)\right]\\
&\times\left[({\bq}^2+\tau')({\bq}^2+\frac{\bar{\tau}}{1+\rho})\right]^{-1}
\end{split}\end{equation}
The above expression can be cast in a form suitable to perform the 
$\bq$-integrals:
\begin{equation}\begin{split}
\delta\Gamma^{(1,0)}=&-\frac 12\lambda\tilde{g}\int\frac{\dd^d q}{(2\pi)^d}\\&
+\frac{\lambda\Psi}{4((1+\rho)\tau'-\bar{\tau})}\left[\tilde{g}(\rho g-2\sigma
f)\tau'-2f^2\bar{g}\Psi \right]\int\frac{\dd^d 
q}{(2\pi)^d}\frac{1}{\bq^2+\tau'}\\&+
\frac{\lambda\Psi}{4(1+\rho)((1+\rho)\tau'-\bar{\tau})}\Bigg[2\tilde{g}\sigma
f\bar{\tau}-(1+\rho)g\tilde{g}\frac{\sigma f}{\rho}\Psi\\&
\phantom{\,-\,}-4\rho f^2 
\bar{\tau}+2
g(1+\rho)f^2\Psi\Bigg]\int\frac{\dd^d 
q}{(2\pi)^d}\frac{1}{\bq^2+\frac{\bar{\tau}}{1+\rho}}
\end{split}
\end{equation}
We use dimensional regularization to compute the momentum integrals:
\begin{equation}
\int\frac{\dd^d q}{(2\pi)^d}\frac{1}{\bq^2+\tau'}=-\frac{2}{\eps}\tau^{\prime
1-\eps/2}A_\eps,\;\;
\int\frac{\dd^d
q}{(2\pi)^d}\frac{1}{\bq^2+\frac{\bar{\tau}}{1+\rho}}=-\frac{2}{\eps}\left(\frac
{
\bar{\tau}}{1+\rho}\right)^{
1-\eps/2}A_\eps
\end{equation}
In terms of renomalized quantities the equation of state has the form
\begin{equation}\begin{split}
h_{\sR}=&\Psi_{\sR}\left\{
\tau_{\sR}+\frac{\bar{g}_{\sR}}{2}\Psi_{\sR}
+\frac{1}{4(1+\rho_{\sR})}
\left[((1+\rho_{\sR})u-4v-2w)
\frac{\bar{\tau}_{\sR}}{1+\rho_{\sR}}
\ln\frac{\mu^{-2}\bar{\tau}_{\sR}}{1+\rho_{\sR}}
\right.\right.\\
&\left. \left.
+\frac{(\rho_{\sR}u-2w)\tau^\prime_{\sR}-2v\bar{g}_{\sR}\Psi_{\sR}}
{\bar{\tau}_{\sR}-(1+\rho_{\sR})\tau^\prime_{\sR}}
\left(
\bar{\tau}_{\sR}
\ln\frac{\mu^{-2}\bar{\tau}_{\sR}}{1+\rho_{\sR}}-
(1+\rho_{\sR})\tau^\prime_{\sR}\ln(\mu^{-2}\tau_{\sR}^\prime)
\right)
\right]\right\}
\end{split}\end{equation}
which is Eq.~(\ref{equationdetat}).
\newpage
\section*{Figures}

\vfill\newpage
\parbox[t]{\textwidth}{\begin{picture}(0,0)%
\epsfig{file=art2fig4.pstex}%
\end{picture}%
\setlength{\unitlength}{2368sp}%
\begingroup\makeatletter\ifx\SetFigFont\undefined%
\gdef\SetFigFont#1#2#3#4#5{%
  \reset@font\fontsize{#1}{#2pt}%
  \fontfamily{#3}\fontseries{#4}\fontshape{#5}%
  \selectfont}%
\fi\endgroup%
\begin{picture}(8712,2856)(901,-3961)
\put(4351,-2536){\makebox(0,0)[lb]{\smash{\SetFigFont{7}{8.4}{\rmdefault}{\mddefault}{\updefault}
$\beta=1-\frac{\eps}{32}+{\cal O}(\eps^2)$
}}}
\put(1576,-2536){\makebox(0,0)[lb]{\smash{\SetFigFont{7}{8.4}{\rmdefault}{\mddefault}{\updefault}
$\beta=1$
}}}
\put(1501,-3961){\makebox(0,0)[lb]{\smash{\SetFigFont{7}{8.4}{\rmdefault}{\mddefault}{\updefault}$D_A<D_B$}}}
\put(901,-1261){\makebox(0,0)[lb]{\smash{\SetFigFont{7}{8.4}{\rmdefault}{\mddefault}{\updefault}$\langle\psi(t=+\infty)\rangle$}}}
\put(6526,-3961){\makebox(0,0)[lb]{\smash{\SetFigFont{7}{8.4}{\rmdefault}{\mddefault}{\updefault}$\tau$}}}
\put(7501,-3961){\makebox(0,0)[lb]{\smash{\SetFigFont{7}{8.4}{\rmdefault}{\mddefault}{\updefault}$D_A>D_B$}}}
\put(9526,-3961){\makebox(0,0)[lb]{\smash{\SetFigFont{7}{8.4}{\rmdefault}{\mddefault}{\updefault}$\tau$}}}
\put(4501,-3961){\makebox(0,0)[lb]{\smash{\SetFigFont{7}{8.4}{\rmdefault}{\mddefault}{\updefault}$D_A=D_B$}}}
\put(3601,-3961){\makebox(0,0)[lb]{\smash{\SetFigFont{7}{8.4}{\rmdefault}{\mddefault}{\updefault}$\tau$}}}
\end{picture}
}

\vfill\newpage
\parbox[t]{\textwidth}{\vskip -4cm\hskip -2cm
 \begin{picture}(0,0)%
\epsfig{file=art2fig7.pstex}%
\end{picture}%
\setlength{\unitlength}{0.00058300in}%
\begingroup\makeatletter\ifx\SetFigFont\undefined%
\gdef\SetFigFont#1#2#3#4#5{%
  \reset@font\fontsize{#1}{#2pt}%
  \fontfamily{#3}\fontseries{#4}\fontshape{#5}%
  \selectfont}%
\fi\endgroup%
\begin{picture}(9957,14057)(-11,-13206)
\put(1651,-1861){\makebox(0,0)[lb]{\smash{\SetFigFont{17}{20.4}{\rmdefault}{\mddefault}{\updefault}$\Psi$}}}
\put(8551,-8536){\makebox(0,0)[lb]{\smash{\SetFigFont{17}{20.4}{\rmdefault}{\mddefault}{\updefault}$\rho$}}}
\end{picture}
} 
\vskip -4cm

\vfill\newpage
\parbox[t]{\textwidth}{\vskip -4cm\hskip -2cm \begin{picture}(0,0)%
\epsfig{file=art2fig6.pstex}%
\end{picture}%
\setlength{\unitlength}{0.00058300in}%
\begingroup\makeatletter\ifx\SetFigFont\undefined%
\gdef\SetFigFont#1#2#3#4#5{%
  \reset@font\fontsize{#1}{#2pt}%
  \fontfamily{#3}\fontseries{#4}\fontshape{#5}%
  \selectfont}%
\fi\endgroup%
\begin{picture}(9957,14057)(-11,-13206)
\put(1651,-1936){\makebox(0,0)[lb]{\smash{\SetFigFont{17}{20.4}{\rmdefault}{\mddefault}{\updefault}$\Psi$}}}
\put(8626,-8536){\makebox(0,0)[lb]{\smash{\SetFigFont{17}{20.4}{\rmdefault}{\mddefault}{\updefault}$\rho$}}}
\end{picture}
}
\vskip -4cm

\parbox[t]{\textwidth}
{\begin{picture}(0,0)%
\epsfig{file=art2fig3.pstex}%
\end{picture}%
\setlength{\unitlength}{0.00058300in}%
\begingroup\makeatletter\ifx\SetFigFont\undefined%
\gdef\SetFigFont#1#2#3#4#5{%
  \reset@font\fontsize{#1}{#2pt}%
  \fontfamily{#3}\fontseries{#4}\fontshape{#5}%
  \selectfont}%
\fi\endgroup%
\begin{picture}(7295,4395)(1118,-4000)
\put(4801,239){\makebox(0,0)[lb]{\smash{\SetFigFont{8}{9.6}{\rmdefault}{\mddefault}{\updefault}$\bar{g}\tilde{g}$}}}
\put(2401,-3961){\makebox(0,0)[lb]{\smash{\SetFigFont{8}{9.6}{\rmdefault}{\mddefault}{\updefault}$D_A<D_B$}}}
\put(6301,-3961){\makebox(0,0)[lb]{\smash{\SetFigFont{8}{9.6}{\rmdefault}{\mddefault}{\updefault}$D_A>D_B$}}}
\put(8401,-3661){\makebox(0,0)[lb]{\smash{\SetFigFont{8}{9.6}{\rmdefault}{\mddefault}{\updefault}$w$}}}
\end{picture}
}
\vfill\newpage

\parbox[t]{\textwidth}{\begin{picture}(0,0)%
\epsfig{file=art2fig1.pstex}%
\end{picture}%
\setlength{\unitlength}{3947sp}%
\begingroup\makeatletter\ifx\SetFigFont\undefined%
\gdef\SetFigFont#1#2#3#4#5{%
  \reset@font\fontsize{#1}{#2pt}%
  \fontfamily{#3}\fontseries{#4}\fontshape{#5}%
  \selectfont}%
\fi\endgroup%
\begin{picture}(5572,7579)(54,-6728)
\put(5626,-3961){\makebox(0,0)[lb]{\smash{\SetFigFont{12}{14.4}{\rmdefault}{\mddefault}{\updefault}
$\Psi$
}}}
\put(211,-1051){\makebox(0,0)[lb]{\smash{\SetFigFont{12}{14.4}{\rmdefault}{\mddefault}{\updefault}
$h$
}}}
\put(5551,-3211){\makebox(0,0)[lb]{\smash{\SetFigFont{12}{14.4}{\rmdefault}{\mddefault}{\updefault}
$\tau<\tau_{spinod}$
}}}
\put(5551,-2326){\makebox(0,0)[lb]{\smash{\SetFigFont{12}{14.4}{\rmdefault}{\mddefault}{\updefault}
$\tau=\tau_{spinod}$
}}}
\put(5551,-1411){\makebox(0,0)[lb]{\smash{\SetFigFont{12}{14.4}{\rmdefault}{\mddefault}{\updefault}
$\tau>\tau_{spinod}$
}}}
\end{picture}
}
\vfill

\parbox[t]{\textwidth}{\vskip -5cm\hskip -2cm\begin{picture}(0,0)%
\epsfig{file=art2fig2.pstex}%
\end{picture}%
\setlength{\unitlength}{3947sp}%
\begingroup\makeatletter\ifx\SetFigFont\undefined%
\gdef\SetFigFont#1#2#3#4#5{%
  \reset@font\fontsize{#1}{#2pt}%
  \fontfamily{#3}\fontseries{#4}\fontshape{#5}%
  \selectfont}%
\fi\endgroup%
\begin{picture}(6072,8228)(-11,-7377)
\put(5476,-5176){\makebox(0,0)[lb]{\smash{\SetFigFont{12}{14.4}{\rmdefault}{\mddefault}{\updefault}
$\tau=\tau_{spinod}$
}}}
\put(6061,-1561){\makebox(0,0)[lb]{\smash{\SetFigFont{12}{14.4}{\rmdefault}{\mddefault}{\updefault}
$\Psi(r_\perp)$
}}}
\put(5896,-4276){\makebox(0,0)[lb]{\smash{\SetFigFont{12}{14.4}{\rmdefault}{\mddefault}{\updefault}
$\tau=\tau_{coex}$
}}}
\put(151,-5401){\makebox(0,0)[lb]{\smash{\SetFigFont{12}{14.4}{\rmdefault}{\mddefault}{\updefault}
$\Psi'(r_\perp)$
}}}
\end{picture}
}
\vskip -4cm

\vfill
\parbox[t]{\textwidth}{\hskip -1.5cm\begin{picture}(0,0)%
\epsfig{file=art2fig5.pstex}%
\end{picture}%
\setlength{\unitlength}{0.00058300in}%
\begingroup\makeatletter\ifx\SetFigFont\undefined%
\gdef\SetFigFont#1#2#3#4#5{%
  \reset@font\fontsize{#1}{#2pt}%
  \fontfamily{#3}\fontseries{#4}\fontshape{#5}%
  \selectfont}%
\fi\endgroup%
\begin{picture}(9237,6570)(376,-6100)
\put(3901,-6061){\makebox(0,0)[lb]{\smash{\SetFigFont{8}{9.6}{\rmdefault}{\mddefault}{\updefault}$\tau<\tau_{coex}$}}}
\put(376,314){\makebox(0,0)[lb]{\smash{\SetFigFont{8}{9.6}{\rmdefault}{\mddefault}{\updefault}$F[\Psi]$}}}
\put(2926,-2461){\makebox(0,0)[lb]{\smash{\SetFigFont{8}{9.6}{\rmdefault}{\mddefault}{\updefault}$\Psi$}}}
\put(601,-6061){\makebox(0,0)[lb]{\smash{\SetFigFont{8}{9.6}{\rmdefault}{\mddefault}{\updefault}$\tau_{coex}$}}}
\put(601,-2461){\makebox(0,0)[lb]{\smash{\SetFigFont{8}{9.6}{\rmdefault}{\mddefault}{\updefault}$\tau>\tau_{spinod}$}}}
\put(3901,-2461){\makebox(0,0)[lb]{\smash{\SetFigFont{8}{9.6}{\rmdefault}{\mddefault}{\updefault}$\tau_{spinod}$}}}
\put(7201,-2461){\makebox(0,0)[lb]{\smash{\SetFigFont{8}{9.6}{\rmdefault}{\mddefault}{\updefault}$\tau_{spinod}>\tau>\tau_{coex}$}}}
\end{picture}
}
\vfill

\end{document}